%% file: paper.tex
\documentclass[
  aps,
  prd,
  reprint,
  onecolumn,
  showpacs,
  nofootinbib,
]{revtex4-1}

\usepackage{microtype}   

\usepackage[UKenglish]{babel}  

\usepackage{siunitx}           
\DeclareSIUnit\year{yr}
\DeclareSIUnit\yr{\year}
\DeclareSIUnit\barn{b}
\DeclareSIUnit\fb{\femto\barn}
\DeclareSIUnit\pb{\pico\barn}

\usepackage{graphicx}   
\usepackage{subcaption} 

\graphicspath{{./images/}}

\usepackage{xcolor}     

\usepackage{tikz}       
\usepackage{pgfplots}   
\usepackage[compat=1.1.0]{tikz-feynman}  
\pgfplotsset{compat=1.14}

\usetikzlibrary{external}
\tikzset{
    external/up to date check=simple,
    external/system call={
        lualatex \tikzexternalcheckshellescape -halt-on-error -interaction=batchmode -jobname="\image" "\texsource" || rm "\image.pdf"},
}

\usepgfplotslibrary{
  colorbrewer,
}

\input{pgfplots.default.tex}

\usepackage{amsmath}     
\usepackage{amssymb}     
\usepackage{mathtools}   
\usepackage{slashed}     
\usepackage{bm}

\usepackage{array}     
\usepackage{booktabs}  

\newcolumntype{C}{>{\(}c<{\)}}
\newcolumntype{L}{>{\(}l<{\)}}
\newcolumntype{R}{>{\(}r<{\)}}

\usepackage{hyperref}  
\usepackage{cleveref}  

\newcommand*\Esix{\relax\ifmmode\mathrm{E}_{6}\else\(\mathrm{E}_{6}\)\fi}
\newcommand*\SO[1]{\relax\ifmmode\mathrm{SO}(#1)\else\(\mathrm{SO}(#1)\)\fi}
\newcommand*\SU[1]{\relax\ifmmode\mathrm{SU}(#1)\else\(\mathrm{SU}(#1)\)\fi}
\newcommand*\Uone[1]{\relax\ifmmode\mathrm{U}(1)_{#1}\else\(\mathrm{U}(1)_{#1}\)\fi}
\newcommand*\Gfive{\relax\ifmmode\mathfrak{G}_{5}\else\(\mathfrak{G}_{5}\)\fi}

\newcommand*\calL{\mathcal{L}}
\newcommand*\calV{\mathcal{V}}

\newcommand*\scL{\textsc{l}}
\newcommand*\scR{\textsc{r}}
\newcommand*\scY{\textsc{y}}
\newcommand*\SM{\textsc{sm}}

\newcommand*\defeq{\vcentcolon=}
\newcommand*\eqdef{=\vcentcolon}

\DeclarePairedDelimiter\abs{\lvert}{\rvert}
\newcommand*\phc{+ \text{h.c.}}

\DeclareMathOperator{\Br}{Br}
\let\Re\relax

\DeclareMathOperator{\Re}{\mathfrak{Re}}

\def\figwidth{0.7}
\def\figheight{0.4}

\begin{document}

\preprint{PREPRINT}

\title{\texorpdfstring{
    Phenomenological Analysis of an \(\mathrm{E}_{6}\)-inspired Seesaw Model
  }{
    Phenomenological Analysis of an E6-inspired Seesaw Model
  }}

\author{Joshua P.~Ellis}
\email{josh@jpellis.me}
\author{Raymond R.~Volkas}
\email{raymondv@unimelb.edu.au}
\affiliation{%
  \textsc{Arc} Centre of Excellence for Particle Physics at the Terascale\\
  School of Physics, The University of Melbourne, Victoria 3010, Australia%
}

\date{\today}


\begin{abstract}
  We analyse the phenomenology of a model of neutrino masses inspired by
  unification into \Esix{} in which the exotic neutrinos can be present at low
  scales.  The model introduces vector-like isosinglet down-type quarks,
  vector-like isodoublet leptons, neutrino singlets and two \(Z'\) bosons.  The
  seesaw mechanism can be achieved with exotic neutrino masses as low as
  \SI{100}{\GeV} and Yukawa couplings of order \num{e-3}.  We find that the
  lightest \(Z'\) boson mass is required to be above \SI{2.8}{\TeV}, the exotic
  quark masses are required to be above \SI{1.3}{\TeV} (\SI{810}{\GeV}) if they
  are collider stable (promptly decaying), and the exotic lepton mass bounds
  remain at the \textsc{lep} value of \SI{102}{\GeV}.  The model also presents a
  type-\textsc{ii} two-Higgs-doublet model (\textsc{2hdm}) along with two heavy
  singlet scalars.  The \textsc{2hdm} naturally has the alignment limit enforced
  thanks to the large vacuum expectation values of the exotic scalars, thereby
  avoiding most constraints.
\end{abstract}

\maketitle

\section{Introduction}
\label{sec:introduction}

The origin of neutrino masses and their unusually small values in comparison to
the other fermions of the Standard Model (\textsc{sm}) are still mysteries in
particle physics.  The seesaw mechanisms provide an elegant way of generating
small masses by introducing either heavy fermions (Type-\textsc{i}
\cite{minkowski77_at_rate_one_out_muon_decay,
  yanagida79_horizo_symmet_mass_neutr, gellmann80_complex_spinor_unified,
  mohapatra80_neutr_mass_spont_parit_noncon} and Type-\textsc{iii}
\cite{foot89_seesaw_neutrin_mass_induc_triplet}) or heavy scalars
(Type-\textsc{ii} \cite{magg80_neutrin_mass_probl_gaug_hier,
  schechter80_neutrin_mass_theories,
  cheng80_neutrin_mass_mixing_oscill,lazarides80_proton_lifetime_fermion_mass,
  wetterich81_neutrin_mass_scale_violat,
  mohapatra80_neutrin_mass_mixing_gauge_model}).  The first of these models is
notoriously difficult to probe experimentally as either the masses of the new
particles are out of the reach of current experiments, or they are too weakly
coupled to the \textsc{sm}.  The type-\textsc{ii} and type-\textsc{iii} models
are more testable because of their gauge interactions, but the purest
incarnation of the seesaw mechanism would still place the new particles beyond
the reach of the \textsc{lhc} because of their generically very large masses.
Variations which bring the new physics into the experimentally testable regime
are therefore of considerable interest.  In this vein,
\citeauthor{cai16_tev_scale_pseud_dirac_seesaw} presented in
Ref.~\cite{cai16_tev_scale_pseud_dirac_seesaw} a seesaw model inspired by
unification into \Esix{}~\cite{guersey76_univer_gauge_theor_model_based_e6,
  Achiman:1978vg,Shafi:1978gg,Barbieri:1981yy} which is realized at scales
testable at the Large Hadron Collider (\textsc{lhc}).  The purpose of this paper
is to explore the rich phenomenology of this model in detail.

In the model, the realization of a \si{\TeV}-scale seesaw mechanism is achieved
thanks to the multiple heavy counterparts of the light \textsc{sm} neutrinos.
In addition, new exotic charged leptons and quarks are introduced in order to
complete the \(\bm{27}\) representation of \Esix{}, and two \(Z'\) bosons arise
from the \Uone{} gauge groups remaining after \Esix{} breaking.  In order to
generate the necessary masses of the exotic fermions, two scalar singlets are
also introduced in this model.

The paper is organized as follows: in \cref{sec:the_model}, we introduce the
particle content of the model in order to realize the seesaw mechanism and show
in \cref{subsec:gauge_unification} how the introduction of additional scalars
can allow the gauge couplings to unify at the grand unified theory
(\textsc{gut}) scale.  In \cref{subsec:decay_of_exotic_quarks}, the issue of
having possibly long-lived coloured particles is addressed and
\cref{subsec:neutrino_masses_and_mixing} explicitly looks at the realization of
the light neutrino masses and the mixing between \textsc{sm} and exotic
neutrinos.  In \cref{sec:constraints}, collider constraints are recast for the
various new particles introduced in this model.


\section{The Model}
\label{sec:the_model}

The model is inspired by grand unification of the \textsc{sm} gauge group into
\Esix{}, which has the subgroup chain
\begin{equation}
  \label{eq:symmetry_breaking}
  \begin{split}
    \Esix
    &\supset \SO{10} \otimes \Uone{\psi} \\
    &\supset \SU{5} \otimes \Uone{\chi} \otimes \Uone{\psi} \eqdef \Gfive.
  \end{split}
\end{equation}
The \textsc{sm} fermions within each generation can transform under the
\(\bm{27}\) irreducible representation (irrep) of \Esix{} which decomposes into
the following irreps of \Gfive{}:
\begin{align}
  \bm{27} &\to (\bm 1)(0, -4) \tag*{\(\in \bm 1_{\SO{10}}\)} \\
  &\quad + (\bm 5)(2, 2) + (\overline{\bm 5})(-2, 2) \tag*{\(\in \bm{10}_{\SO{10}}\)} \\
  &\quad + (\bm 1)(-5, -1) + (\overline{\bm 5})(3, -1) + (\bm{10})(-1, -1).
  \tag*{\(\in \bm{16}_{\SO{10}}\)}
\end{align}

The \textsc{sm} fermions are contained within the last two terms above and will
be denoted by \(\psi_{10}\) and \(\psi_{\overline 5}\).  The singlet of
\(\bm{16}_{\SO{10}}\) is an exotic singlet neutrino \(\psi_{1}\).  The remaining
\(\bm 5\), \(\overline{\bm 5}\) and \(\bm 1\) irreps contain only exotic
particles: \(\bm 1\) corresponds to an additional singlet neutrino \(\chi_{1}\),
while \(\chi_{5} \sim (\bm 5)(2,2)\) and \(\chi_{\overline{5}} \sim
(\overline{\bm 5})(-2,2)\) contain a vector-like isosinglet down-type quark
\(B\) and a vector-like lepton isodoublet \(R = (\nu_{r}, r)\).  The gauge
charges of all the fermions are summarized in
\cref{subtab:quantum_numbers_fermion}.

The Yukawa terms of the \textsc{sm} originate from coupling
\(\chi_{27}\chi_{27}\) to a \(\bm{27}\) irrep of scalars.  The scalar
\(\bm{27}\) decomposes similarly to the fermions and contains two Higgs doublets
in the \(\bm{5}\) and \(\overline{\bm{5}}\) irreps from
\(\bm{10}_{\SO{10}}\). The allowed couplings between these two Higgs doublets
and the \textsc{sm} fermions gives rise to a type-\textsc{ii} two-Higgs-doublet
model (\textsc{2hdm}).

While we have motivated the particle content using representations of \Esix{},
we emphasize that this model is only \emph{inspired} by that unification group,
and thus does not comply with every restriction it would impose.  We do,
however, make contact with \Esix{} whenever appropriate in order to set up a
possible eventual derivation from a complete unified theory.\footnote{To
  understand what may be involved in achieving a full \Esix{} realization, see
  for example Refs.~\cite{Bajc2014,Babu2015}.}

In order to ensure that the exotic fermions in \(\chi_{5, \overline 5}\) are
sufficiently heavy, it will also be assumed that \(\Phi_{1}\) [from the
\(\bm{1}_{\SO{10}}\) -- see \cref{subtab:quantum_numbers_scalar}] gains a
nonzero vacuum expectation value (\textsc{vev}) generating an appropriately
large mass term. The Higgs doublets residing in the \(H_5 \sim (\bm{5})(2,2)\)
and \(H_{\overline{5}} \sim (\overline{\bm{5}})(-2,2)\) quintuplets are, of
course, required to gain electroweak-scale \textsc{vev}s and masses.  All other
scalars from the decomposition of \(\bm{27}\) will be absent in our
\textsc{gut}-inspired theory, as will the color-triplet partners of the Higgs
doublets in \(H_{5,\overline{5}}\).

\begin{table}
  \centering
  \begin{subtable}[t]{0.49\linewidth}
    \centering
    \begin{tabular}{LRRR}
      \toprule
                         & \SU{5}            & \Uone{\chi} & \Uone{\psi}
      \\
      \midrule
      \psi_{10}          & \bm{10}           & -1          & -1 \\
      \psi_{\overline 5} & \overline{\bm{5}} & 3           & -1 \\
      \psi_{1}           & \bm{1}            & -5          & -1 \\
      \midrule
      \chi_{5}           & \bm{5}            & 2           & 2 \\
      \chi_{\overline 5} & \overline{\bm{5}} & -2          & 2 \\
      \chi_{1}           & \bm{1}            & 0           & -4 \\
      \bottomrule
    \end{tabular}
    \caption{Fermionic fields}
    \label{subtab:quantum_numbers_fermion}
  \end{subtable}
  \begin{subtable}[t]{0.49\linewidth}
    \centering
    \begin{tabular}{LRRR}
      \toprule
                            & \SU{5}            & \Uone{\chi} & \Uone{\psi}
      \\
      \midrule
      H_{5}                 & \bm{5}            & 2           & 2 \\
      H_{\overline 5}       & \overline{\bm{5}} & -2          & 2 \\
      \Phi_{1}              & \bm{1}            & 0           & -4 \\
      \midrule
      \midrule
      \Phi_{2}              & \bm{1}            & 5           & 5 \\
      \Phi_{3}              & \bm{1}            & 5           & -3 \\
      \bottomrule
    \end{tabular}
    \caption{Scalar fields}
    \label{subtab:quantum_numbers_scalar}
  \end{subtable}
  \caption[Transformation properties of the matter content under \Gfive{}.]{
    Transformation properties of the matter content under \Gfive{}. The
    \textsc{sm} fermions are contained within \(\psi_{10}\) and
    \(\psi_{\overline 5}\).  The \(H_{5,\overline 5}\) contain the two Higgs
    doublets from the type-\textsc{ii} \textsc{2hdm}.  Other than
    \(\Phi_{2,3}\), all particles originate from the \(\bm{27}\) irrep of
    \Esix{}.  It is possible to obtain \(\Phi_{2}\) and \(\Phi_{3}\) from the
    \(\bm{351}\) and \(\bm{78}\) irreps of \Esix{} respectively.}
  \label{tab:quantum_numbers}
\end{table}

At the \Gfive{} scale, the Yukawa couplings are
\begin{equation}
  \label{eq:g5_yukawa}
  \begin{split}
    \MoveEqLeft \calL_{\text{Yuk}}
    = y_{u} H_{5} \psi_{10} \psi_{10} + y_{d} H_{\overline 5} \psi_{\overline 5} \psi_{10} \\
    &+ y_{xu} H_{5} \chi_{\overline 5} \chi_{1} + y_{xd} H_{\overline 5} \chi_{5} \chi_{1} \\
    &+ y_{\nu} H_{5} \psi_{\overline 5} \psi_{1} + y_{1} \Phi_{1} \chi_{5} \chi_{\overline 5}
    \phc,
  \end{split}
\end{equation}
and below the \Gfive{} scale, the last term splits
\begin{equation}
  \label{eq:y1_low_scale}
  y_{1} \Phi_{1} \chi_{5} \chi_{\overline 5} \to y_{1d} \Phi_{1} \overline B B + y_{1\ell} \Phi_{1} \overline R R.
\end{equation}
In these equations, \Esix{} and \SU{5} restrictions on the Yukawa coupling
constants are \emph{not} imposed.

In order to produce seesaw-suppressed neutrino masses, it turns out that an
additional Yukawa interaction,
\begin{equation}
  \label{eq:phi2_yukawa}
  y_{2} \Phi_{2} \chi_{1} \psi_{1} \phc,
\end{equation}
must be introduced.  The required additional scalar \(\Phi_{2}\) must transform
as \((\bm{1})(5, 5)\) under \Gfive{} and can originate from the \(\bm{351}\)
irrep of \Esix{}.  The scalars introduced and their transformation properties
under \Gfive{} are presented in \cref{subtab:quantum_numbers_scalar}.  (The role
of \(\Phi_{3}\) will be discussed in \cref{subsec:decay_of_exotic_quarks}.)

\subsection{Gauge Unification}
\label{subsec:gauge_unification}

Grand unified theories (\textsc{gut}s) are generally motivated in the context of
supersymmetry because the contribution of the extra Higgs doublet and the
superpartners to the renormalization group running of the gauge coupling
constants ensures that they obtain a common value at a phenomenologically
acceptable \textsc{gut} scale.  Furthermore, the naturalness problem posed by
the large hierarchy between the \textsc{gut} and electroweak scales can be
avoided.  The model studied in this paper is, however, nonsupersymmetric and
thus the particle content has to be adjusted in order that unification can still
be achieved.  Despite the model being only inspired by \Esix{} grand
unification, we pause to analyse how acceptable gauge coupling constant
unification could in principle arise.\footnote{In our \textsc{gut}-inspired
  effective scenario, the potential naturalness problem is not too severe
  because all of the new particles have masses of at most several TeV.  Of
  course, a \textsc{gut} completion would have a problem.}  We will be
considering a direct breaking of \Esix{} directly to the \textsc{sm} gauge group
at the high scale with no intermediate scale.

With only the vector-like fermions \(\chi_{5,\overline{5}}\), the doublets from
\(H_{5,\overline 5}\), and \(\Phi_{1}\) and \(\Phi_{2}\) contributing, the
running of the gauge couplings do not unite exactly, with \(g_{2} = g_{3}
\approx \num{0.6}\) at a scale of \SI{e16}{\GeV} while the hypercharge coupling
constant \(g_{1} \approx \num{0.7}\).  In order to achieve unification, or
near-exact unification, additional scalars that do not fill out complete \SU{5}
representations can be introduced at some intermediate energy scale.  We demand
full unification of the \textsc{sm} gauge coupling constants and the coupling
constants \(g_{4}\) and \(g_{5}\) of \Uone{\chi} and \Uone{\psi},
respectively. All the \Uone{} coupling constants are normalized as if they are
embedded in \Esix{}.

One way that unification can be achieved is by invoking additional \SU{2}
doublets.  As we are already using the \(\Phi_{2}\) from the \(\bm{351}\) irrep
of \Esix{}, we may assume that additional scalars from this multiplet (which do
not gain \textsc{vev}s) also survive to lower scales so that they contribute to
the renormalization group running.  An example that achieves unification without
the introduction of colored states uses scalars in the representations:
\begin{equation}
  \label{eq:extra_doublets}
  \begin{aligned}
    (\bm{1}, \bm{2})&(3, 3, -1), &
    (\bm{1}, \bm{2})&(3, -2, 2), \\
    (\bm{1}, \bm{2})&(-3, 2, 2), &
    (\bm{1}, \bm{2})&(-3, 7, -1), \\
    (\bm{1}, \bm{2})&(-3, -3, 5),
  \end{aligned}
\end{equation}
all being introduced at \SI{20}{\TeV}, a scale chosen for the sake of
definiteness.  This option, however, results in a low unification scale of
\SI{e14}{\GeV} which would be at odds with bounds from proton decay in the
context of a hypothetical \textsc{gut} completion.

\begin{figure}
  \centering
  \tikzsetnextfilename{unification}
  \begin{tikzpicture}
    \begin{semilogxaxis}[
        width=\figwidth\linewidth,
        height=\figheight\linewidth,
        xlabel={Scale [\si{\GeV}]},
        ylabel={Coupling},
        legend columns=3,
        enlarge x limits=false,
        cycle list/Dark2-5,
        table/col sep=comma,
        table/x=scale,
      ]

      \begin{scope}[/pgfplots/restrict x to domain=0:38.5093]
        \addplot+ table [y=g1c] {paper_data/rge.csv};
        \addlegendentry{\Uone{\scY}}
        \addplot+ table [y=g2c] {paper_data/rge.csv};
        \addlegendentry{\SU{2}}
        \addplot+ table [y=g3c] {paper_data/rge.csv};
        \addlegendentry{\SU{3}}
        \addplot+ table [y=g4c] {paper_data/rge.csv};
        \addlegendentry{\Uone{\chi}}
        \addplot+ table [y=g5c] {paper_data/rge.csv};
        \addlegendentry{\Uone{\psi}}
      \end{scope}

      \begin{scope}[/pgfplots/restrict x to domain=0:32.637]
        \addplot+ [dashed] table [y=g1] {paper_data/rge.csv};
        \addplot+ [dashed] table [y=g2] {paper_data/rge.csv};
        \addplot+ [dashed] table [y=g3] {paper_data/rge.csv};
        \addplot+ [dashed] table [y=g4] {paper_data/rge.csv};
        \addplot+ [dashed] table [y=g5] {paper_data/rge.csv};
      \end{scope}
    \end{semilogxaxis}
  \end{tikzpicture}
  \caption{Running of the gauge coupling constants with energy scale calculated
    at one-loop order.  The solid curves show the running when additional
    scalars charged under \SU{3} are considered as per
    \cref{eq:extra_coloured_doublets}, while the dashed curves consider only the
    \SU{3} singlets of \cref{eq:extra_doublets}.  The additional states used to
    achieve unification are all introduced at \SI{20}{\TeV}.}
  \label{fig:gauge_running}
\end{figure}
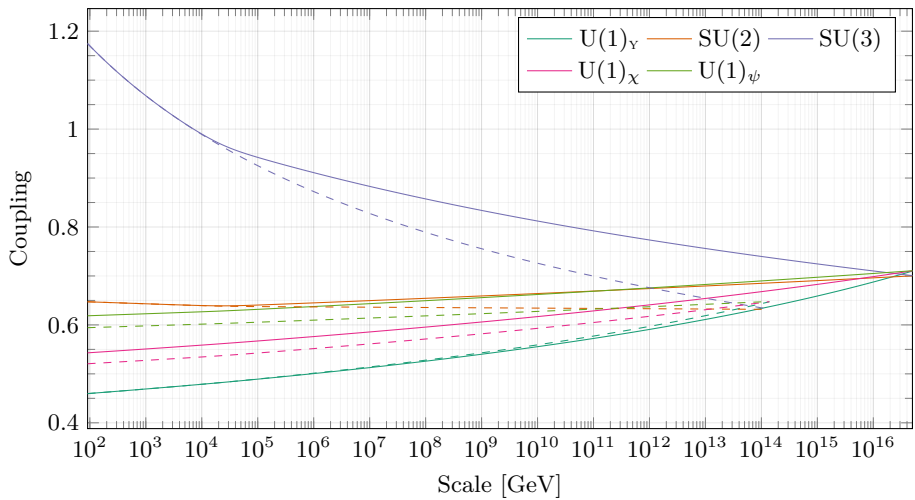

This issue can be rectified by altering the running of the \SU{3} coupling
constant such that the unification occurs at a higher scale.  In particular, the
combination of
\begin{equation}
  \label{eq:extra_coloured_doublets}
  \begin{aligned}
    (\overline{\bm{3}}, \bm{2})&(1, 1, 5), &
    (\bm{3}, \bm{2})&(-1, -1, -1), \\
    (\bm{1}, \bm{2})&(3, -2, 2), &
    (\overline{\bm{6}}, \bm{2})&(-1, -1, -1),
  \end{aligned}
\end{equation}
all introduced at \SI{20}{\TeV} results in a very good agreement between all
five gauge coupling constants above \SI{e16}{\GeV}.  As before, these additional
scalars can all originate from the \(\bm{351}\) irrep of \Esix{}.  The solutions
of the regnormalization group equations (\textsc{rge}s) in both cases are
depicted in \cref{fig:gauge_running}.  These were calculated using SARAH
\cite{staub14_sarah} at one-loop order, and then numerically evaluated using
SPheno \cite{porod12_spheno} and FlexibleSUSY
\cite{athron15_flexib_spect_gener_gener_super_model}.  Note that despite the
differences at high scales, the low-scale phenomenology remains mostly unchanged
with the coupling constants for \Uone{\chi} and \Uone{\psi} evolving to nearly
the same low-energy values in both cases.  Contributions arising at two-loop
order to the \textsc{rge} running were also investigated, but found to not alter
the unification significantly.

Note that \cref{fig:gauge_running} does not appear to exhibit exact coupling
constant unification, but that is merely because of some simplifying
assumptions.  We have taken \Esix{} to break to the \textsc{sm} in one step,
rather than through a cascade involving \SO{10} and \SU{5} at intermediate
scales. Also, threshold effects have been neglected. Removing these
simplifications will allow full unification to occur, given that the convergence
of the coupling constants is already quite precise in our simplified analysis.

In any \textsc{gut} completion, proton decay would be mediated at least through
the \(X\) and \(Y\) gauge bosons, resulting in a proton lifetime at tree level
of
\begin{equation}
  \label{eq:proton_lifetime}
  \tau_{p} \sim  \frac{1}{g^{4}} \frac{m_{X,Y}^{4}}{m_{p}^{5}}.
\end{equation}
The \(X\) boson mediates the decay \(p \to e^{+} \pi^{0}\), while the \(Y\)
boson can mediate both \(p \to e^{+} \pi^{0}\) and \(p \to \overline \nu
K^{+}\). The \(X\) and \(Y\) bosons gain their masses from the scalar that is
responsible for breaking \SU{5} into the \textsc{sm} gauge groups, leading to
\(m_{X,Y} = g v_{\textsc{gut}} / \sqrt{2}\).  Using an estimate from
Ref.~\cite{langacker81_grand_unified_theor_proton_decay} of the proton lifetime
that is more accurate than \cref{eq:proton_lifetime}, one obtains
\begin{equation}
  \label{eq:proton_lifetime_num}
  \tau_{p} \approx
  \begin{cases}
    \SI{5.2e26}{\year} & v_{\textsc{gut}} \sim \SI{e14}{\GeV} \\
    \SI{3.3e37}{\year} & v_{\textsc{gut}} \sim \SI{e16}{\GeV}
  \end{cases},
\end{equation}
using our value of the gauge coupling constant \(g\) at the unification
scale. At present, Super-Kamiokande has found that \(\tau / \Br(p \to e^{+}
\pi^{0}) > \SI{8.2e33}{\year}\) \cite{asakura15_searc_proton_decay_mode_p}, and
KamLAND has found that \(\tau (p \to \overline \nu K^{+}) > \SI{5.4e32}{\year}\)
\cite{nishino09_searc_proton_decay_via_p}. Thus the scenario of
\cref{eq:extra_coloured_doublets} is clearly phenomenologically allowed.

\subsection{Decay of Exotic Quarks}
\label{subsec:decay_of_exotic_quarks}

The exotic quarks in \(\chi_{5,\overline 5}\) pose a problem in this model as
they cannot decay at tree level.  In a \textsc{gut} completion of this model,
the decay can take place through the coloured components of \(H_{5,\overline
  5}\) quintuplets, though the decay width will remain small as the coloured
Higgs components are required to be extremely heavy in order to evade
constraints from proton decay.  Having long-lived coloured exotic particles is
problematic as it interferes with nucleosynthesis and as a result, the lifetime
of the exotic quarks needs to be less than \SI{0.1}{\second}
\cite{kawasaki05_big_bang_nucleos_hadron_decay,reno88_primor_nucleos}.

At dimension five (\textsc{d5}), the only gauge-invariant term contributing to
the decay of the exotic quarks is
\begin{equation}
  \label{eq:d5_terms_no_phi3}
  \calL_{\textsc{d5}} \supset \frac{1}{\Lambda} \Phi_{1}^{\dagger} \Phi_{2}^{\dagger}\chi_{5} \psi_{\overline 5}
\end{equation}
which introduces mixing between the \textsc{sm} and exotic down-type quarks. We
will be assuming that the mixing is confined within each generation so that it
is sufficient to consider the \(2 \times 2\) mixing:
\begin{subequations}
  \begin{align}
    \label{eq:b-B_mass_matrix}
    \calL_{d-B} &=
    \begin{pmatrix}
      \overline d_{\scL} & \overline B_{\scL}
    \end{pmatrix}
    \begin{pmatrix}
      m_{d} & 0 \\
      \frac{v_{1} v_{2}}{2 \Lambda} & m_{B}
    \end{pmatrix}
    \begin{pmatrix}
      d_{\scR} \\
      B_{\scR}
    \end{pmatrix} \phc \\
    \label{eq:b-B_diagonal_mass_matrix}
    &=
    \begin{pmatrix}
      \overline d'_{\scL} & \overline B'_{\scL}
    \end{pmatrix}
    U_{\scL}
    \begin{pmatrix}
      m_{d}' & 0 \\
      0 & m_{B}'
    \end{pmatrix}
    U_{\scR}^{\dagger}
    \begin{pmatrix}
      d'_{\scR} \\
      B'_{\scR}
    \end{pmatrix} \phc,
  \end{align}
\end{subequations}
where the primed fields denote the mass eigenstates, and \(m_{d}\) and \(m_{B}\)
are the original masses generated from the Yukawa interactions with the Higgs
and \(\Phi_{1}\) respectively.  This mixing introduces a small mass correction
to the two original masses:
\begin{subequations}
  \label{eq:b_mass_corrections}
  \begin{align}
    m_{d}' - m_{d} &= -\frac{1}{2} \left( \frac{v_{1} v_{2}}{2 \Lambda} \right)^{2} \frac{m_{d}}{m_{B}^{2}}, \\
    m_{B}' - m_{B} &= \frac{1}{2} \left( \frac{v_{1} v_{2}}{2 \Lambda} \right)^{2} \frac{1}{m_{B}}.
  \end{align}
\end{subequations}
For the parameters investigated, the resulting correction to the down-type
\textsc{sm} quarks is about 1 part in \num{e10} at most.  We are taking
\(\Lambda\) to be the \textsc{gut} scale, \SI{e16}{\GeV}.

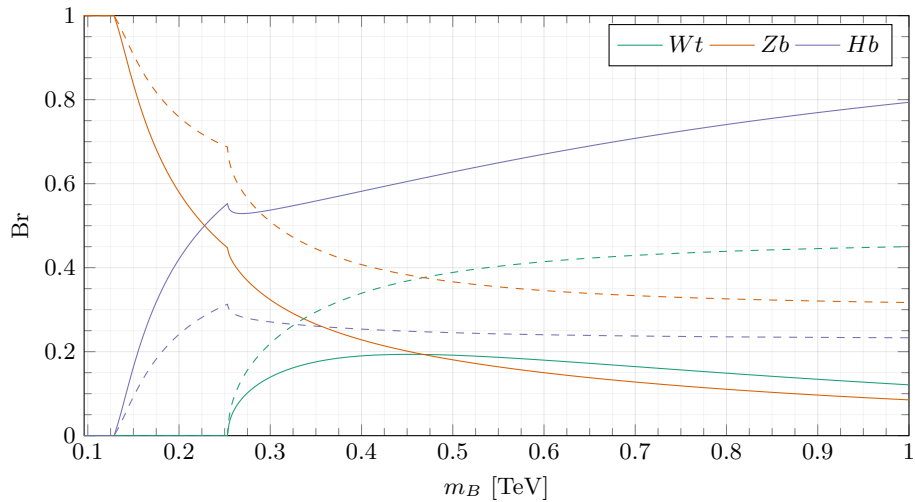
\begin{figure}
  \centering
  \tikzsetnextfilename{quark_br}
  \begin{tikzpicture}
    \begin{axis}[
        width=\figwidth\linewidth,
        height=\figheight\linewidth,
        xlabel={\(m_{B}\) [\si{\TeV}]},
        scaled x ticks=manual:{}{\pgfmathparse{#1/1000}},
        enlarge x limits=false,
        restrict x to domain=0:1000,
        enlarge y limits=false,
        ylabel=\(\Br\),
        legend columns=-1,
        table/col sep=comma,
        table/x=mB,
        cycle list/Dark2-3,
      ]

      \addplot+ table [y=Wt-v3] {paper_data/exotic_quark_br.csv};
      \addlegendentry{{\(Wt\)}}
      \addplot+ table [y=Zb-v3] {paper_data/exotic_quark_br.csv};
      \addlegendentry{{\(Zb\)}}
      \addplot+ table [y=Hb-v3] {paper_data/exotic_quark_br.csv};
      \addlegendentry{{\(Hb\)}}

      \addplot+[dashed, Dark2-A, forget plot] table [y=Wt-nov3] {paper_data/exotic_quark_br.csv};
      \addlegendentry{{\(W^{-} t\)}}
      \addplot+[dashed, Dark2-B, forget plot] table [y=Zb-nov3] {paper_data/exotic_quark_br.csv};
      \addlegendentry{{\(Zb\)}}
      \addplot+[dashed, Dark2-C, forget plot] table [y=Hb-nov3] {paper_data/exotic_quark_br.csv};
      \addlegendentry{{\(Hb\)}}
    \end{axis}
  \end{tikzpicture}
  \caption{Decay branching ratios of the third generation exotic quark into
    third generation \textsc{sm} quarks.  In the solid lines, the \textsc{vev}
    of \(\Phi_{3}\) is taken to be \SI{e9}{\GeV} so that the exotic quarks decay
    promptly.  The dashed lines indicate the branching ratios in the absence of
    \(\Phi_{3}\).  Decays mediated by virtual particles are not included.}
  \label{fig:exotic_quark_br}
\end{figure}

The \(d\)--\(B\) quark mixing introduces new terms allowing for the exotic
quarks to decay to \(Wu\), \(Zd\) and \(Hd\) and resulting in a decay width
\begin{equation}
  \label{eq:d5_decay_no_phi3}
  \Gamma_{\textsc{d5}}(B) \sim \left( \frac{v_{1} v_{2}}{2 \Lambda} \right)^{2} \frac{1}{m_{B}}.
\end{equation}
The tree-level partial widths are described in the appendix, along with the
relevant couplings to the \(W\) and \(Z\) gauge bosons, and the Higgs boson.  In
collider searches, we will be primarily interested in final states involving
third-generation \textsc{sm} quarks, in which case the branching fractions as
functions of the exotic quark mass are plotted in \cref{fig:exotic_quark_br}
assuming the benchmark configuration of \textsc{vev}s described in
\cref{subsec:exotic_Z_bosons} (\(v_{1} = \SI{23}{\TeV}\), \(v_{2} =
\SI{25}{\TeV}\)).

\begin{figure}
  \centering
  \tikzsetnextfilename{quark_bbn}
  \begin{tikzpicture}
    \begin{loglogaxis}[
        width=\figwidth\linewidth,
        height=\figheight\linewidth,
        log ticks with fixed point,
        xlabel={\(m_{B}\) [\si{\TeV}]},
        xtick={200, 500, 1000, 2000, 5000},
        minor xtick={300, 400, 600, 700, 800, 900, 1500, 3000, 4000},
        xticklabels={0.2, 0.5, 1.0, 2.0, 5.0},
        enlarge x limits=false,
        ylabel={\(\tau_{B}\) [\si{\s}]},
        ytick={0.01, 0.02, 0.05, 0.1, 0.2},
        minor ytick={0.03, 0.04, 0.06, 0.07, 0.08, 0.09, 0.15},
        table/col sep=comma,
        table/x=mB,
        cycle list/Dark2-5,
      ]

      \addplot+ [y=lifetime] table {paper_data/exotic_quark_lifetime.csv};
      \draw [red] (200, 0.1) -- (5000, 0.1)
        node [pos=0.1, above] {\textsc{bbn}};
    \end{loglogaxis}
  \end{tikzpicture}
  \caption{Lifetime of the exotic \(B\) quarks for the benchmark \textsc{vev}
    configuration from \cref{subsec:exotic_Z_bosons}.  In red is the upper bound
    on the lifetime of the exotic quarks due to \textsc{bbn}
    \cite{kawasaki05_big_bang_nucleos_hadron_decay, reno88_primor_nucleos}.}
  \label{fig:B_lifetime}
\end{figure}
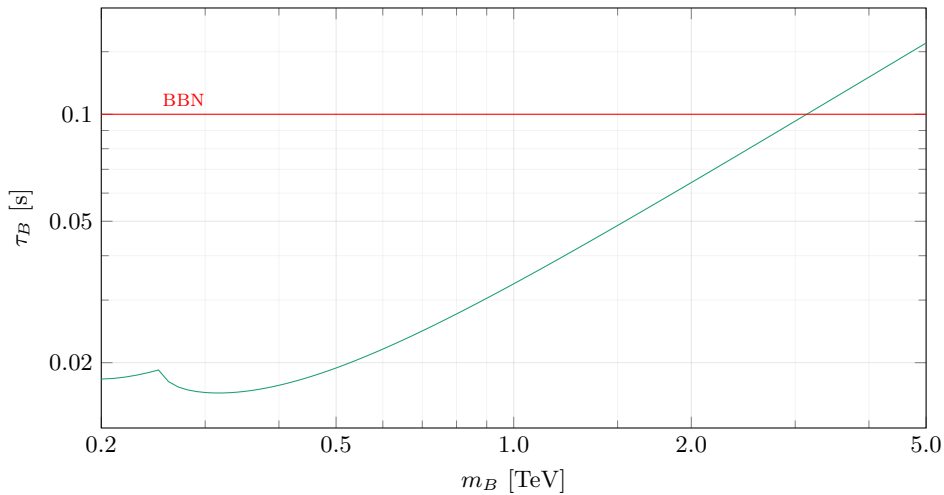

In order to satisfy the Big Bang nucleosynthesis (\textsc{bbn}) constraint,
\(v_{1,2}\) must be at least in the low \si{\TeV} range; however due to the
decreased \(d\)--\(B\) mixing with larger mass separation, there is an upper
bound on the exotic quark masses for a given \(v_{1}\) and \(v_{2}\).  In
particular, the benchmark configuration of \textsc{vev}s requires that \(m_{B} <
\SI{3}{\TeV}\) as can be seen in \cref{fig:B_lifetime}.  The remaining range of
allowed masses results in collider-stable exotic quarks.

In Ref.~\cite{cai16_tev_scale_pseud_dirac_seesaw}, an additional scalar
\(\Phi_{3}\) was introduced in order to facilitate the decay of the exotic
quarks in case the \textsc{vev}s of \(\Phi_{1}\) and \(\Phi_{2}\) were
insufficient to satisfy the \textsc{bbn} constraints.  The quantum numbers of
\(\Phi_{3}\) are shown in \cref{subtab:quantum_numbers_scalar}, and it
introduces the following additional \textsc{d5} terms that facilitate the decay
of the exotic quark:
\begin{equation}
  \label{eq:d5_terms_with_phi3}
  \calL_{\textsc{d5}} \supset \frac{1}{\Lambda} \left[ \Phi_{1} \Phi_{3}^{\dagger} \chi_{5} \psi_{\overline 5} + \Phi_{3} H_{\overline 5} \chi_{5} \psi_{10} \right].
\end{equation}
The first term introduces another contribution to the off-diagonal entry in the
mass matrix in \cref{eq:b-B_mass_matrix}, while the second term introduces
direct coupling between the Higgs, bottom quark and exotic quark after
\(\Phi_{3}\) gains a \textsc{vev}.  As this last term is not suppressed by the
decreased mixing that accompanies larger \(d\)--\(B\) mass separations, the
decay \(B \to Hd\) becomes the dominant decay mode for heavy exotic quark masses
and scales according to
\begin{equation}
  \label{eq:d5_decay_with_phi3}
  \Gamma_{\textsc{d5}}(B \to Hd) \sim \left( \frac{v_{3}}{\Lambda} \right)^{2} m_{B}.
\end{equation}
The branching fraction to \(B \to Hd\) reaches \SI{90}{\percent} at \(m_{B} =
\SI{1.7}{\TeV}\) and \SI{95}{\percent} at \SI{2.6}{\TeV} for the benchmark
configuration of \textsc{vev}s and \(v_{3} = \SI{e9}{\GeV}\).

\subsection{Neutrino Masses and Mixing}
\label{subsec:neutrino_masses_and_mixing}

Below the electroweak scale and remaining in the one-generation approximation,
there are five neutral fermions which will mix and generate the seesaw
mechanism.  Their mass matrix in the \((\psi_{\overline 5}, \psi_{1},
\chi_{\overline 5}, \chi_{5}, \chi_{1})\) basis is
\begin{equation}
  \label{eq:neutrino_mass_matrix}
  \mathrm{M}_{\nu}
  = \frac{1}{\sqrt{2}} \begin{pmatrix}
    0              & -y_{\nu} v_{u} & 0               & 0               & 0 \\
    -y_{\nu} v_{u} & 0              & 0               & 0               & y_{2} v_{2} \\
    0              & 0              & 0               & y_{1\ell} v_{1} & - y_{xu} v_{u} \\
    0              & 0              & y_{1\ell} v_{1} & 0               & - y_{xd} v_{d} \\
    0              & y_{2} v_{2}    & - y_{xu} v_{u}  & - y_{xd} v_{d}  & 0
  \end{pmatrix}.
\end{equation}
The different signs within the mass matrix arise due to the differences in the
\SU{2} contractions.  For example, the \textsc{sm} neutrino mass term with
\SU{2} indices explicitly written is
\begin{equation}
  y_{\nu} H_{u}^{\alpha} \overline L^{\beta} \psi_{1} \varepsilon_{\alpha\beta}
  \equiv y_{\nu} \left[ H_{u}^{-} \overline e_{\scL} - H_{u}^{0} \overline \nu_{\scL} \right] \psi_{1}.
\end{equation}
In the case of the \(y_{1\ell}\) and \(y_{2}\) Yukawa terms, the sign is
positive because all fields are \SU{2} singlets.

With \(v_{1}\) and \(v_{2}\) both much larger than \(v_{u,d}\), one mass
eigenstate will be very light, two will be \(\sim v_{1}\) and two will be \(\sim
v_{2}\).  The latter two pairs form pseudo-Dirac fermions with small mass
splittings.  In the scenario where \(y_{1\ell} v_{1}\) and \(y_{2} v_{2}\) are
nondegenerate and both larger than all other terms in the mass matrix
(\cref{eq:neutrino_mass_matrix}), the mass eigenstates of all the neutrinos are
well approximated by:\footnote{The neutrino mass spectrum per family bears some
  resemblance to that of the inverse seesaw mechanism
  \cite{Mohapatra:1986bd,GonzalezGarcia:1988rw}, in that there is one very light
  Majorana eigenstate and, in our case, two very massive pseudo-Dirac pairs,
  compared to one massive pseudo-Dirac state for the inverse seesaw. However,
  there is no analogue in our case of the very small explicit
  lepton-number-violating parameter typical of the inverse seesaw mechanism.}

\begin{subequations}
  \label{eq:neutrino_masses}
  \begin{align}
    \label{eq:neutrino_mass_light}
    m_{\nu} &\approx \frac{\sqrt{2} y_{\nu}^{2} y_{xd} y_{xu}}{y_{1\ell} y_{2}^{2}} \frac{v_{d} v_{u}^{3}}{v_{1} v_{2}^{2}}, \\
    \label{eq:neutrino_mass_1}
    m_{N_{1}} &\approx \frac{y_{1\ell} v_{1}}{\sqrt{2}},  \\
    \label{eq:neutrino_mass_2}
    m_{N_{2}} &\approx \frac{y_{2} v_{2}}{\sqrt{2}}.
  \end{align}
\end{subequations}
Alternatively the mass scale of the \textsc{sm} neutrinos can be expressed as
\begin{equation}
  \label{eq:sm_neutrino}
  m_{\nu} \approx \frac{y_{\nu}^{2} y_{xd} y_{xu}}{2} \frac{v_{\SM}^{4}}{m_{N_{1}} m_{N_{2}}^{2}} \cos \beta \sin^{3} \beta,
\end{equation}
where \(\tan\beta \defeq v_u/v_d\) and \(m_{N_{1,2}}\) are taken to be exactly
as given in \cref{eq:neutrino_mass_1,eq:neutrino_mass_2}.  It should be
re-iterated that, in general, the exact mass eigenstates ought to be calculated
from the matrix itself as the above expressions are only true in certain
limiting cases.  For example, in scenarios where \(y_{1\ell} v_{1} \sim v_{2}
v_{2}\) the above approximations no longer hold.  Additionally, certain small
contributions have been omitted in the expressions above as they are generally
insignificant.  For example, there are higher-order terms in \(m_{\nu}\) which
are suppressed by larger factors of \(v_{1,2}\), and the two pseudo-Dirac masses
in \(m_{N_{1,2}}\) have small contributions proportional to \(v_{u,d}\).

The masses were calculated by setting the Yukawa couplings in
\cref{eq:neutrino_mass_matrix} at the \textsc{gut} scale and running them down
to low scales using one-loop \textsc{rge}s.  Due to the smallness of \textsc{sm}
neutrino masses, corrections occurring at two-loop order have a possibility of
introducing sizeable corrections to the light neutrino masses, though in our
case no such issues were encountered: the masses of the \textsc{sm} neutrinos
received only a minor correction.

Current constraints from Planck place an upper bound of \SI{290}{\meV} on the
sum of light neutrino masses \cite{plank2015_resul_cosmo_param}, while current
best fits on neutrino observables
\cite{capozzi17_global_const_absol_neutr_masses_their_order} require that the
sum of light neutrino masses be at least \SI{60}{\meV} and \SI{100}{\meV} for
the normal and inverted hierarchies respectively.

\begin{figure}
  \centering
  \tikzsetnextfilename{yukawa_couplings}
  \begin{tikzpicture}
    \begin{axis}[
        width=\figwidth\linewidth,
        height=\figheight\linewidth,
        samples=51,
        domain=100:1000,
        xlabel={\(y_{1\ell} v_{1} / \sqrt{2}\) [\si{\GeV}]},
        y domain=100:1000,
        ylabel={\(y_{2} v_{2} / \sqrt{2}\) [\si{\GeV}]},
        point meta rel=per plot,
        view={0}{90},
        colorbar horizontal,
        colorbar style={
          at={(0.5, 1.03)},
          anchor=south,
          xticklabel pos=upper,
          xlabel=\(\sqrt[4]{y_{xd} y_{xu} y_{\nu}^{2}}\),
        },
        table/col sep=comma,
        table/x=m1,
        table/y=m2,
      ]

      \addplot3[
        surf,
        shader=interp,
        colorbar source,
      ] table [z=y] {paper_data/neutrino_yukawa.csv};

      \addplot3[
        contour gnuplot={
          levels={1e-3,2e-3,3e-3,4e-3},
          label distance=4cm,
          draw color={white},
          contour label style={
            /pgf/number format/fixed,
            /pgf/number format/precision=3,
            every node/.style={
              white,
              anchor=north,
            }
          }
        },
      ] table [z=y] {paper_data/neutrino_yukawa.csv};

      \addplot3[
        opacity=0.5,
        contour gnuplot={
          levels={1.5e-3,2.5e-3,3.5e-3},
          labels=false,
          draw color={white},
        },
      ] table [z=y] {paper_data/neutrino_yukawa.csv};
    \end{axis}
  \end{tikzpicture}
  \caption{Values of the geometric mean of the four Yukawa couplings which will
    generate a neutrino mass of \SI{50}{\meV} assuming that \(\tan \beta = 10\),
    as a function of the two exotic neutrino masses.  The seesaw mechanism can
    easily be achieved with relatively light exotic neutrinos and Yukawa
    couplings similar to those found in the \textsc{sm}.  The white contour
    lines show specific values of \(\sqrt[4]{y_{xd} y_{xu} y_{\nu}^2}\).}
  \label{fig:neutrino_yukawa_contours}
\end{figure}
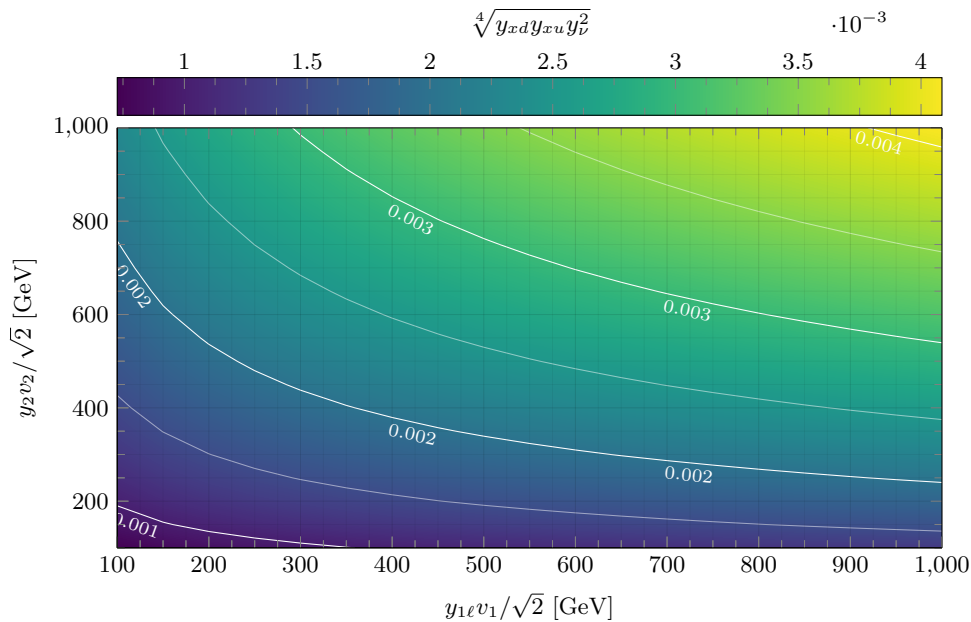

This model easily achieves light neutrino masses on the order of \SI{10}{\meV}
while avoiding having extremely small Yukawa couplings.  For example, with
\(\tan \beta = 10\), the values of the product \(y_{xu} y_{xd} y_{\nu}^{2}\)
which will generate a neutrino mass of \SI{50}{\meV} are shown in
\cref{fig:neutrino_yukawa_contours}.  Even in the case where exotic neutrinos
masses are \(\sim \SI{100}{\GeV}\), we can realize the desired lightness of the
\textsc{sm} neutrinos provided that \(y_{xu} y_{xd} y_{\nu}^{2} \approx
(\num{e-3})^{4}\).

The \textsc{sm} neutrinos mix primarily with \(\psi_{1}\) and \(\chi_{1}\), and
only negligibly with the neutral components of \(\chi_{5, \overline 5}\) even if
these are much lighter than \(\psi_{1}\) and \(\chi_{1}\) (though still assuming
they are significantly heavier than the \textsc{sm} neutrinos) making it
sufficient to consider the simplified neutrino mass matrix,
\begin{equation}
  \label{eq:simplified_neutrino_mass_matrix}
  \frac{1}{\sqrt{2}} \begin{pmatrix}
    0 & - y_{\nu} v_{u} & 0 \\
    - y_{\nu} v_{u} & 0 & y_{2} v_{2} \\
    0 & y_{2} v_{2} & 0
  \end{pmatrix},
\end{equation}
when evaluating the mixing between the \textsc{sm} neutrinos and exotic
neutrinos.\footnote{Note that this simplified matrix is sufficient for analysing
  the \emph{mixing} only, but does not describe the \textsc{sm} light neutrino
  mass eigenvalues (in fact, they are zero in this approximation).}  In
particular, the mixing of \textsc{sm} neutrinos with exotic neutrinos is largely
independent of \(y_{1\ell} v_{1}\) despite the \textsc{sm} neutrino mass being
proportional to \(1 / y_{1\ell} v_{1}\).  This results in the neutrino mixing
and neutrino masses not being as strongly linked as in the conventional seesaw
mechanism, thereby allowing this model to have simultaneously large mixing and
quite small mass separations.  Another consequence is that the mixing of the
\textsc{sm} neutrino need not be primarily with the lightest exotic neutrino.

This mixing with exotic neutrinos causes the Pontecorvo--Maki--Nakagawa--Sakata
(\textsc{pmns}) matrix to deviate from unitarity which has repercussions for a
number of lepton flavour and electroweak observables.  Following the notation of
Ref.~\cite{Fernandez-Martinez:2016lgt}, this deviation from unitarity can be
encapsulated in \(\eta\) defined by:
\begin{equation}
  N = (I - \eta) U_{\textsc{pmns}},
\end{equation}
where \(N\) is the matrix describing the mixing between the light neutrino mass eigenstates and the \textsc{sm} charged leptons via \(W\) interactions.  In the one-generation approximation, the deviation from unitarity is
\begin{equation}
  \label{eq:eta_approx}
  2 \eta_{\alpha \alpha} = \left[ 1 + \left(\frac{y_{2} v_{2}}{y_{\nu} v_{\SM} \sin \beta} \right)^2 \right]^{-1}.
\end{equation}

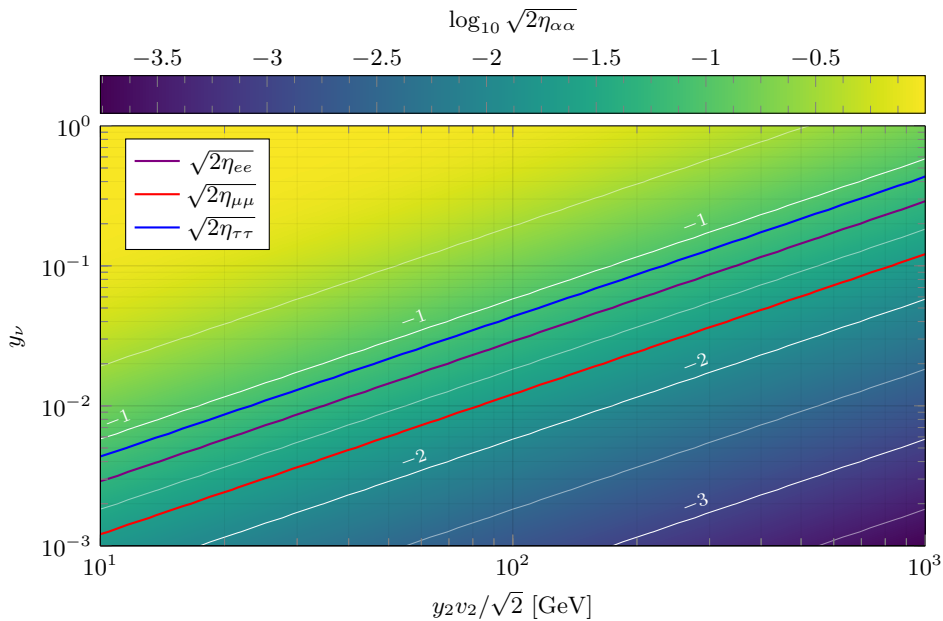
\begin{figure}
  \centering
  \tikzsetnextfilename{pmns_bound}
  \begin{tikzpicture}
    \begin{loglogaxis}[
        width=\figwidth\linewidth,
        height=\figheight\linewidth,
        domain=50:1000,
        xlabel={\(y_{2} v_{2} / \sqrt{2}\) [\si{\GeV}]},
        domain y=1e-3:1,
        ylabel={\(y_{\nu}\)},
        view={0}{90},
        colorbar horizontal,
        colorbar style={
          at={(0.5, 1.03)},
          anchor=south,
          xticklabel pos=upper,
          xlabel={\(\log_{10} \sqrt{2 \eta_{\alpha\alpha}}\)},
        },
        legend pos=north west,
        table/col sep=comma,
        table/x=m2,
        table/y=yv,
      ]

      \addplot3[
        forget plot,
        surf,
        shader=interp,
        colorbar source,
      ] table [z=log10(sqrt(2eta))] {paper_data/eta.csv};

      \addplot3[
        forget plot,
        contour gnuplot={
          levels={-4, -3, -2, -1, 0},
          draw color={white},
          label distance=4cm,
          contour label style={
            every node/.style={
              white,
              anchor=south,
            }
          }
        },
      ] table [z=log10(sqrt(2eta))] {paper_data/eta.csv};
      \addplot3[
        forget plot,
        opacity=0.5,
        contour gnuplot={
          levels={-4.5, -3.5, -2.5, -1.5, -0.5},
          draw color={white},
          labels=false,
        },
      ] table [z=log10(sqrt(2eta))] {paper_data/eta.csv};

      \addplot3[
        red!50!blue,
        thick,
        contour gnuplot={
          levels={-1.3010}, 
          draw color={red!50!blue},
          labels=false,
        },
      ] table [z=log10(sqrt(2eta))] {paper_data/eta.csv};
      \addlegendentry{{\(\sqrt{2 \eta_{ee}}\)}}

      \addplot3[
        red,
        thick,
        contour gnuplot={
          levels={-1.6778}, 
          draw color={red},
          labels=false,
        },
      ] table [z=log10(sqrt(2eta))] {paper_data/eta.csv};
      \addlegendentry{{\(\sqrt{2 \eta_{\mu\mu}}\)}}

      \addplot3[
        blue,
        thick,
        contour gnuplot={
          levels={-1.1249}, 
          draw color={blue},
          labels=false,
        },
      ] table [z=log10(sqrt(2eta))] {paper_data/eta.csv};
      \addlegendentry{{\(\sqrt{2 \eta_{\tau\tau}}\)}}
    \end{loglogaxis}
  \end{tikzpicture}
  \caption{ Bound on \(y_{\nu}\) and \(y_{2} v_{2}\) due to the deviation from
    unitarity of the \textsc{pmns} matrix based on a global fit to lepton
    flavour and electroweak data by \cite{Fernandez-Martinez:2016lgt}.  The
    region of parameter space above the red line is excluded.  The bound takes
    the limit \(\tan \beta \to \infty\) and assumes that the neutral components
    of \(\chi_{5, \overline 5}\) are significantly heavier than the \textsc{sm}
    neutrinos (though they need not be heavier than \(\psi_{1}\) and
    \(\chi_{1}\)).  The white contour lines show specific values of \(\sqrt{2
      \eta_{\alpha\alpha}}\).  }
  \label{fig:pmns_unitarity_bound}
\end{figure}

A global fit to lepton flavour and electroweak data places \(2 \sigma\) upper
bounds on \(\sqrt{2 \eta_{ee}}\), \(\sqrt{2 \eta_{\mu\mu}}\) and \(\sqrt{2
  \eta_{\tau\tau}}\) of \num{0.050}, \num{0.021} and \num{0.075} respectively
\cite{Fernandez-Martinez:2016lgt}.  The allowed parameter space in \(y_{2}
v_{2}\)--\(y_{\nu}\) is shown in \cref{fig:pmns_unitarity_bound} and restricts
\(y_{\nu}\) to be at most \num{e-2} if both \(\psi_{1}\) and \(\chi_{1}\) are
around \SI{100}{\GeV}.

\section{Constraints}
\label{sec:constraints}

\subsection{\texorpdfstring{\(Z'\) Bosons}{Z' Bosons}}
\label{subsec:exotic_Z_bosons}

In addition to the photon and the \textsc{sm} \(Z\) boson, the model features
two new massive neutral gauge bosons originating from the exotic \Uone{} gauge
groups.  The most significant contribution to the exotic gauge boson masses
originates from the large nonzero \textsc{vev}s of \(\Phi_{1,2}\), though as the
two Higgs doublets are both charged under these new \Uone{} gauge groups, some
tree-level mixing between the \(Z'\) bosons and the \textsc{sm} \(Z\) boson is
introduced.\footnote{Note that kinetic mixing also exists but is small and will
  be neglected.  At any high \textsc{gut} scale, there can be no kinetic mixing
  between the \Uone{} gauge groups, but it will be generated at lower scales
  through radiative corrections.  In this model, the result is small with
  kinetic-mixing coefficients of order \(\num{e-3}\) for each pair of \Uone{}
  gauge groups.}  The matrix of squared masses generated by the symmetry
breaking is
\begin{widetext}
  \begin{equation}
    \label{eq:gauge_boson_mass_matrix}
    \mathrm{M}_{Z} = \frac{1}{4}\begin{pmatrix}
      g_{1}^{2} v_{\SM}^{2}                           & -g_{1} g_{2} v_{\SM}^{2}                         & \frac{g_{1} g_{4}}{\sqrt{5}} v_{\SM}^{2}                                   & -\frac{g_{1} g_{5}}{3} v_{\SM}^{2} \cos 2 \beta \\
      -g_{1} g_{2} v_{\SM}^{2}                        & g_{2} v_{\SM}^{2}                                & - \frac{g_{2} g_{4}}{\sqrt{5}} v_{\SM}^{2}                                 & \frac{1}{3} g_{2} g_{5} v_{\SM}^{2} \cos 2 \beta \\
      \frac{g_{1} g_{4}}{\sqrt{5}} v_{\SM}^{2}        & - \frac{g_{2} g_{4}}{\sqrt{5}} v_{\SM}^{2}       & \frac{g_{4}^{2}}{20} (25 v_{2}^{2} + 4 v_{\SM}^{2})                        & \frac{g_{4} g_{5}}{12 \sqrt{5}}(25 v_{2}^{2} - 4 v_{\SM}^{2} \cos 2 \beta) \\
      -\frac{g_{1} g_{5}}{3} v_{\SM}^{2} \cos 2 \beta & \frac{1}{3} g_{2} g_{5} v_{\SM}^{2} \cos 2 \beta & \frac{g_{4} g_{5}}{12 \sqrt{5}}(25 v_{2}^{2} - 4 v_{\SM}^{2} \cos 2 \beta) & \frac{g_{5}^{2}}{36} (16 v_{1}^{2} + 25 v_{2}^{2} + 4 v_{\SM}^{2})
    \end{pmatrix}
  \end{equation}
\end{widetext}

The mixing between the \textsc{sm} and exotic bosons introduces a modification
to the \(Z\) pole mass and couplings to fermions which have both been measured
very precisely, with the \(Z\) pole mass measurements significantly constraining
the \(Z\)--\(Z'\) mixing \cite{Langacker2009}.  At tree level, the correction to
the \(Z\) pole mass in this model is
\begin{equation}
  \label{eq:Z_pole_mass_modification}
  \frac{m_{Z,\SM}^{2} - m_{Z}^{2}}{m_{Z,\SM}^{2}} \approx \frac{4 v_{\SM}^{2}}{25 v_{2}^{2}} + \frac{v_{\SM}^{2}}{v_{1}^{2}} \frac{1}{(1 + \tan^{2} \beta)^{2}},
\end{equation}
where \(m_{Z,\SM}^{2}\) is the \textsc{sm} tree-level \(Z\) boson mass,
\(m_{Z}^{2}\) is the tree-level mass in this model, and we are assuming
\(v_{\SM} \ll v_{1,2}\).  The \(Z\) boson pole mass has been measured accurately
at \textsc{lep} with the best fit being \SI{91.1875+-0.0021}{\GeV}
\cite{06_precis_elect_measur_z_reson}.  In order that the shift to the \(Z\)
pole be such that the \(Z\) pole remain within \(1 \sigma\) of the experimental
value, we require that \(v_{2}\) be larger than \SI{14.5}{\TeV} (taking \(\tan
\beta \to \infty\)).

The shift in the \(Z\) pole mass also introduces a shift in the electroweak
\(\rho\) parameter,
\begin{equation}
  \rho \defeq \frac{m_{W}^{2}}{m_{Z}^2 \cos^{2} \theta_{\textsc{W}}},
\end{equation}
where \(\theta_{\textsc{W}} \defeq \arctan(g_{2} / g_{1})\).  The best fit for
this parameter was determined by \textsc{lep} to be \num{1.0050+-0.0010}
\cite{06_precis_elect_measur_z_reson}, and just as with the pole mass, requiring
that the shift in \(\rho\) not exceed the standard deviation of the measured value
results in a lower bound on \(v_{2}\) of \SI{3.1}{\TeV}.

If we demand that the gauge couplings unify at the \textsc{gut} scale, this
fixes the interaction of the \(Z'\) bosons with \(\Phi_{1,2}\) and the other
fermions and consequently the masses of both \(Z'\) bosons are only determined
by the \textsc{vev}s of \(\Phi_{1,2}\).  Additionally, the decays of the \(Z'\)
bosons are similarly fixed and depend only on the kinematics of the decays (and
consequently the exotic Yukawa couplings).

\begin{figure}
  \centering
  \tikzsetnextfilename{zprime_bound}
  \begin{tikzpicture}
    \begin{axis}[
        width=\figwidth\linewidth,
        height=\figheight\linewidth,
        xlabel={\(v_{1}\) [\si{\TeV}]},
        scaled x ticks=manual:{}{\pgfmathparse{#1/1000}},
        ylabel={\(v_{2}\) [\si{\TeV}]},
        scaled y ticks=manual:{}{\pgfmathparse{#1/1000}},
        view={0}{90},
        point meta rel=per plot,
        colorbar horizontal,
        colorbar style={
          at={(0.5, 1.03)},
          anchor=south,
          xticklabel pos=upper,
          xlabel={\(m_{Z'}\) [\si{\TeV}]},
          scaled y ticks=manual:{}{\pgfmathparse{##1/1000}},
        },
        unbounded coords=jump,
        table/col sep=comma,
        table/x=v1,
        table/y=v2,
      ]

      \addplot3[
        surf,
        shader=interp,
        colorbar source,
      ] table [z=mass] {paper_data/zp_bounds.csv};
      \addplot3[
        contour gnuplot={
          levels={1000,2000,3000,4000},
          draw color={white},
          label distance=10cm,
          contour label style={
            every node/.style={
              white,
              anchor=north west,
            }
          }
        },
      ] table [z=mass] {paper_data/zp_bounds.csv};

      \addplot3[
        thick,
        contour gnuplot={
          levels={0.5},
          draw color={red},
          labels=false,
        },
      ] table [z=large-y] {paper_data/zp_bounds.csv};

      \addplot3[
        thick,
        dashed,
        contour gnuplot={
          levels={0.5},
          draw color={red},
          labels=false,
        },
      ] table [z=small-y] {paper_data/zp_bounds.csv};

      \node [fill=red, shape=diamond, inner sep=2pt] at (23000, 25000) {};

      \draw [orange] (\pgfkeysvalueof{/pgfplots/xmin}, 14500) -- (\pgfkeysvalueof{/pgfplots/xmax}, 14500);
    \end{axis}
  \end{tikzpicture}
  \caption{Lower bounds on the \textsc{vev}s of \(\Phi_{1}\) and
    \(\Phi_{2}\). The orange line indicates the lower bound on \(v_{2}\) due to
    modification to the \(Z\) pole mass (taking \(\tan \beta \to \infty\)) and
    the red lines show the bounds from the \(pp \to Z' \to \ell\ell\) channel
    (\(\ell = e,~\mu\)) based on the analysis in
    Ref.~\cite{atlas17_dilepton}---lighter \(Z'\) masses being excluded.  The
    red solid and dashed lines show the exclusion with \(y_{1d} = y_{1\ell} =
    y_{2} = 0.5\) and \(y_{1d} = y_{1\ell} = y_{2} = \num{5e-3}\) at the
    \textsc{gut} scale making the exotic fermions heavy and light, respectively.
    The white contours indicate specific values of \(m_{Z'}\).  The point marked
    by the diamond indicates the benchmark configuration of \textsc{vev}s
    investigated in this model.}
  \label{fig:zp_bounds}
\end{figure}
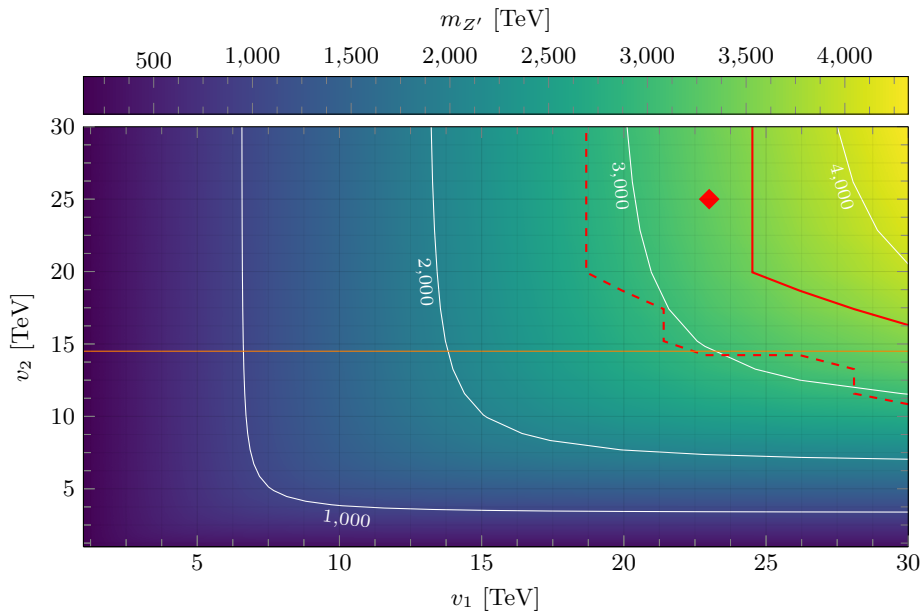

Depending on the exact masses of the exotic fermions, the lighter \(Z'\) boson
mass is restricted to be above \SIrange{2.8}{3.4}{\TeV} as shown in
\cref{fig:zp_bounds}.\footnote{The production and decay of the \(Z'\) was
  calculated within MadGraph \cite{alwall14_autom_comput_tree_level_next} in
  conjunction with Pythia \cite{sjoestrand15_introd_to_pythia}.  The analyses
  were recast using CheckMATE \cite{drees15_checkmate} which builds upon the
  software and algorithms in Refs.~\cite{favereau14_delph,
    cacciari12_fastj_user_manual, cacciari08_anti_kt_jet_clust_algor,
    read02_presen_searc_resul}.}  This limit on the \textsc{vev} configurations
also implies that the heavier \(Z'\) boson mass be at least \SI{7.7}{\TeV}.  For
the remainder of this paper, we will consider the benchmark point \(v_{1} =
\SI{23}{\TeV}\) and \(v_{2} = \SI{25}{\TeV}\) for definiteness.  For this
configuration, the masses of the two \(Z'\) bosons are \SIlist{3.4;10.3}{\TeV}.
In the future, the completion of \textsc{lhc} \SI{14}{\TeV} runs could provide
enough data to exclude the \(Z'\) boson mass up to \SI{6}{\TeV}, and prospects
at a \SI{100}{\TeV} collider could place bounds as high as \SI{30}{\TeV} on the
\(Z'\) masses \cite{Arcadi17}.

As the \textsc{vev}s are actually quite large, this will generally result in
large masses for \(\Phi_{1}\) and \(\Phi_{2}\) unless the quartic coupling
constants in the scalar potential are tuned to achieve light
masses. Additionally, neither scalar can be easily produced at the \textsc{lhc}
as they couple to neither gluons nor the \textsc{sm} quarks at tree level.  In
the case of \(\Phi_{1}\), it can couple to gluons through a loop of exotic
quarks; however, the effective \(\Phi_{1} gg\) coupling is generally
insignificant as either the exotic masses are too heavy and suppress the loop
factor, or the Yukawa coupling is too small.

\subsection{Exotic Down Quarks}
\label{subsec:exotic_down_quarks}

The exotic quarks introduced in this model can be abundantly produced at the
\textsc{lhc} thanks to their couplings to gluons and provide one of the main
phenomenological windows into this model.  Their detection however depends
primarily on their decay mode which, as discussed in
\cref{subsec:decay_of_exotic_quarks}, will be collider-stable in the absence of
\(\Phi_{3}\).

Specifically, with the benchmark \textsc{vev} configuration, the lifetime of the
exotic quarks is \(\sim\) \SI{20}{\milli\second} for \(m_{B} \sim\)
\SI{500}{\GeV}.  With the introduction of \(\Phi_{3}\), the exotic quarks can
decay promptly provided \(v_{3} \gtrsim \SI{e9}{\GeV}\).

\begin{figure}
  \centering
  \tikzsetnextfilename{quark_lhc}
  \begin{tikzpicture}
    \begin{semilogyaxis}[
        width=\figwidth\linewidth,
        height=\figheight\linewidth,
        xlabel={\(m_{B}\) [\si{\TeV}]},
        scaled x ticks=manual:{}{\pgfmathparse{#1/1000}},
        enlarge x limits=false,
        axis x line*=bottom,
        ylabel={\(\sigma_{\SI{13}{\TeV}}(pp \to B \overline{B})\) [\si{\pb}]},
        minor ytick={
          1e-3, 2e-3, 3e-3, 4e-3, 5e-3, 6e-3, 7e-3, 8e-3, 9e-3,
          1e-1, 2e-1, 3e-1, 4e-1, 5e-1, 6e-1, 7e-1, 8e-1, 9e-1,
          1e1, 2e1, 3e1, 4e1, 5e1, 6e1, 7e1, 8e1, 9e1
        },
        table/col sep=comma,
        table/x=mass,
        cycle list/Dark2-5,
      ]

      \addplot+ table [y=13tev-xsec] {paper_data/exotic_quark.csv};

      \def\ymax{300}
      \def\ymin{1e-4}
      \def\arrowlength{0.8cm}

      \draw [|->, red, thick] (1347.7718, 0.0110) -- ++(\arrowlength, 0cm)
        node [pos=1.1, anchor=west] {\textsc{Cms}};

      \draw [|->, red, thick] (1299.14166, 0.0146) -- ++(\arrowlength, 0cm)
        node [pos=1.1, anchor=south] {\textsc{Atlas}};

      \node [red, anchor=north east, align=left] at (1300, 0.0110) {Collider \\ stable};

      \draw [|->, blue, thick] (804.155033, 0.4058) -- ++(\arrowlength, 0cm)
        node [pos=1.1, anchor=west] {\textsc{Cms}};

      \draw [|->, blue, thick] (644.032812, 1.5807) -- ++(\arrowlength, 0cm)
        node [pos=1.1, anchor=west] {\textsc{Atlas}};

      \node [blue, anchor=north east, align=left] at (700, 1) {Prompt \\ decay};
    \end{semilogyaxis}

    \begin{semilogyaxis}[
        width=\figwidth\linewidth,
        height=\figheight\linewidth,
        xlabel={\(y_{1d}\)},
        enlarge x limits=false,
        axis x line*=top,
        xticklabel style={
          /pgf/number format/fixed,
          /pgf/number format/precision=2
        },
        axis y line=none,
        grid=none,
        table/col sep=comma,
        table/x=y1d,
      ]
      \addplot [draw=none] table [y=13tev-xsec] {paper_data/exotic_quark.csv};
    \end{semilogyaxis}
  \end{tikzpicture}
  \caption{Lower bounds on exotic quark masses.  The Yukawa coupling on the top
    axis assumes \(v_{1} = \SI{23}{\TeV}\).  In red are the bounds on
    collider-stable exotic quarks from \textsc{Atlas}
    \cite{16_searc_heavy_long_lived_charg} and \textsc{Cms}
    \cite{cms16_search_long_lived}, and in blue are the bounds on
    promptly decaying exotic quarks from \textsc{Atlas}
    \cite{atlas17_searc_pair_produc_heavy_vector} and \textsc{Cms}
    \cite{cms16_searc_pair_produc_vector_proton}.  The green line shows the
    predicted production cross section of the exotic quarks at the \textsc{lhc}
    at a centre-of-mass energy of \SI{13}{\TeV}, with the multiplicity from the
    three generations taken into account.}
  \label{fig:exotic_quarks}
\end{figure}
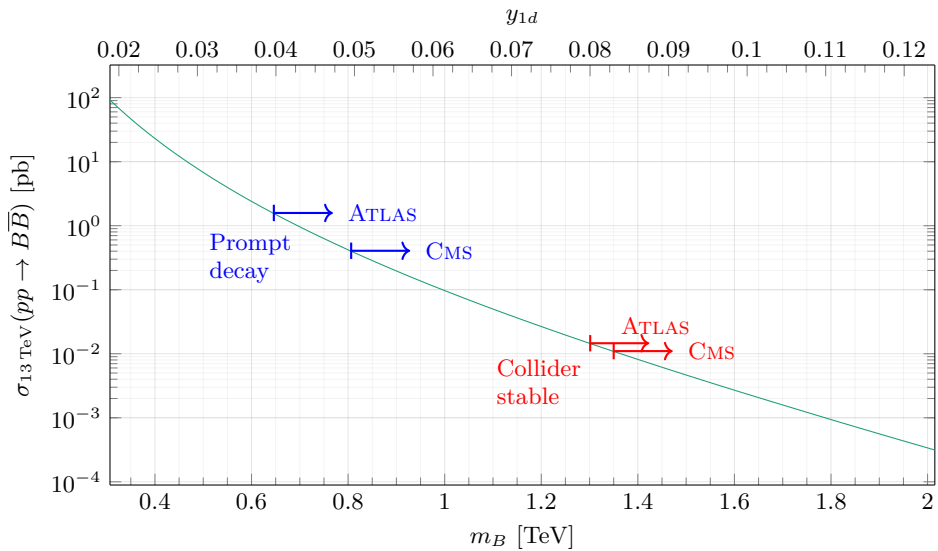

In the scenario where \(\Phi_{3}\) is absent, the exotic quarks hadronize and
form colorless hadrons analogous to \(R\)-hadrons after being created.  As the
hadrons traverse the detector, they leave tracks with large energy losses due to
ionization which have been searched for by both \textsc{Atlas}
\cite{aaboud16_searc_metas_heavy_charg_partic, 16_searc_heavy_long_lived_charg}
and \textsc{Cms} \cite{cms16_search_long_lived}.  As we are assuming for
simplicity that all the exotic quarks have the same mass, the cross section is
enhanced by a factor of 3 leading to slightly more stringent bounds on the quark
mass as shown in \cref{fig:exotic_quarks}.

If the scalar \(\Phi_{3}\) is introduced (despite not being explicitly needed to
satisfy the nucleosynthesis constraint), the exotic quarks will decay promptly
provided \(v_{3} \gtrsim \SI{e9}{\GeV}\).  In this case, searches by
\textsc{Atlas} and \textsc{Cms} have nearly exclusively searched for decays into
third-generation \textsc{sm} quarks as the heavy \textsc{sm} quarks provide a
way to distinguish the exotic quark decays from other background events.  The
resulting limits on exotic quarks depend primarily on the three decay modes:
\begin{align}
  \nonumber
  B \to W t, && B \to Z b, && B \to H b.
\end{align}
Although the \textsc{d5} terms allowing for the exotic quarks to decay could
have complex flavour couplings, we will assume that the three generations of
exotic quarks decay into \textsc{sm} fermions of their corresponding
generations.  The branching ratios to the three final states above are shown
shown in \cref{fig:exotic_quark_br} where the enhancement of \(\Br(B \to Hb)\)
due to the second term in \cref{eq:d5_terms_with_phi3} is evident.  The explicit
partial widths are listed in the appendix.

The most stringent constraints come from a search by \textsc{Cms} in
Ref.~\cite{cms16_searc_pair_produc_vector_proton} which has been particularly
sensitive to decays involving a Higgs, ultimately excluding masses of the third
generation below \SI{810}{\GeV}.  A more recent analysis by \textsc{Cms} with
data collected at \(\sqrt{s} = \SI{13}{\TeV}\) has failed to improve the bounds
on the exotic down-type quarks \cite{cms17_searc_pair_produc_vector_like}.

Searches by \textsc{Atlas} have generally been focusing on final states
involving a boosted \(W\) boson resulting in weaker constraints on the exotic
quarks presented in this model.  Their most recent analysis in
Ref.~\cite{atlas17_searc_pair_produc_heavy_vector} restricted the limits of the
third generation of exotic quarks to \SI{650}{\GeV}.  \textsc{Atlas} has also
conducted a search with light \textsc{sm} quarks in the final states
\cite{atlas15_search_for_pair_heavy_quark}, but this failed to place any
constraints on the exotic quarks in this model as the search was insensitive to
\(B \to Hq\).

\subsection{Exotic Leptons \& Neutrinos}
\label{subsec:exotic_leptons_and_neutrinos}

In the conventional seesaw mechanisms, the new leptons and neutrinos are
typically well out of the reach of present-day experiments.  In some
circumstances, this can be alleviated by modifying the textures of the Dirac
(and Majorana) mass matrices allowing these models to be experimentally probed
more easily \cite{Deppisch:2015qwa}.  In this \Esix{}-inspired model, the exotic
leptons can have electroweak-scale masses while still achieving the desired
suppression of the light neutrino masses.  This makes it in principle possible
to produce them at the \textsc{lhc}, though in most cases the production
cross section is too small for these states to be observed over the backgrounds.

Searches at \textsc{lep} have limited the masses of the exotic charged leptons
to be above \SI{102}{\GeV} \cite{achard01_searc_heavy_neutr_charg_lepton}.  The
same analysis also restricted the masses of exotic neutrinos that decay into a
\(W\) to also be above \SI{102}{\GeV}.  Of the four exotic neutrinos per
generation, only one pair of gauge eigenstates are charged under \SU{2}, but
through mixing, all mass eigenstates ultimately decay into a \(W\) boson
(provided it is kinematically allowed) and a charged lepton.

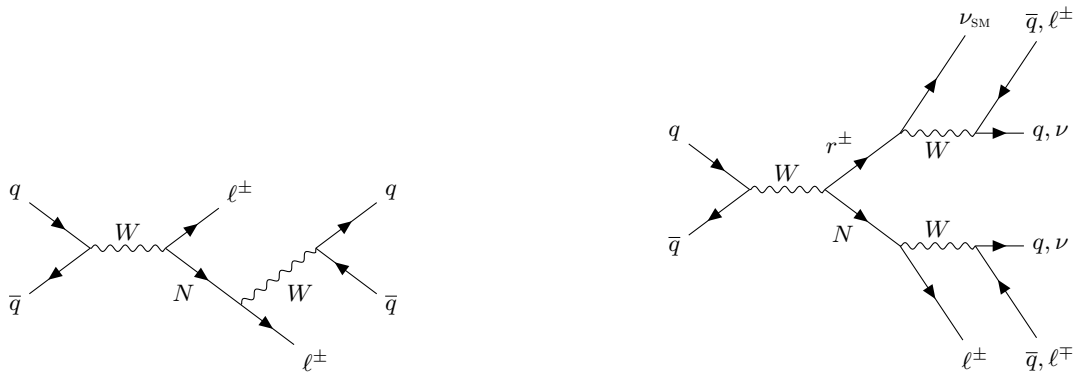
\begin{figure}
  \centering
  \begin{subfigure}[t]{0.49\linewidth}
    \tikzsetnextfilename{qq_Nl}
    \feynmandiagram [small, horizontal=w1 to w2, layered layout] {
      q1 [particle=\(q\)] -- [fermion] w1,
      q2 [particle=\(\overline q\)] -- [anti fermion] w1,
      w1 -- [boson, edge label=\(W\)] w2,
      w2 -- [fermion] l1 [particle=\(\ell^{\pm}\)],
      w2 -- [fermion, edge label'=\(N\)] n,
      n -- [boson, edge label'=\(W\)] wn,
      wn -- [fermion] qf1 [particle={\(q\)}] ,
      wn -- [anti fermion] qf2 [particle={\(\overline q\)}],
      n -- [fermion] l2 [particle=\(\ell^{\pm}\)],
    };
    \caption{Topology used in searches for heavy neutrinos.  This can result in
      same-sign dilepton of different flavours making it more easily
      identifiable.}
    \label{subfig:neutrino_search_topology}
  \end{subfigure}
  \begin{subfigure}[t]{0.49\linewidth}
  \tikzsetnextfilename{qq_Nr}
    \feynmandiagram [small, horizontal=w1 to w2, layered layout] {
      q1 [particle=\(q\)] -- [fermion] w1,
      q2 [particle=\(\overline q\)] -- [anti fermion] w1,
      w1 -- [boson, edge label=\(W\)] w2,
      w2 -- [fermion, edge label=\(r^{\pm}\)] l1,
      l1 -- [boson, edge label'=\(W\)] wl,
      l1 -- [fermion] nu [particle=\(\nu_{\SM}\)],
      wl -- [anti fermion] ql2 [particle={\(\overline q, \ell^{\pm}\)}],
      wl -- [fermion] ql1 [particle={\(q, \nu\)}],
      w2 -- [fermion, edge label'=\(N\)] n,
      n -- [boson, edge label=\(W\)] wn,
      wn -- [fermion] qn1 [particle={\(q, \nu\)}] ,
      wn -- [anti fermion] qn2 [particle={\(\overline q, \ell^{\mp}\)}],
      n -- [fermion] l2 [particle=\(\ell^{\pm}\)],
    };
    \caption{Most likely topology to use in searches for the exotic lepton
      doublet.}
    \label{subfig:exotic_lepton_search_topology}
  \end{subfigure}
  \caption{
    Topologies involved in searches for the exotic leptons.}
  \label{fig:lepton_topologies}
\end{figure}

Searches for heavy neutrinos have been conducted by both \textsc{Atlas}
\cite{aad15_searc_heavy_major_neutr_with} and \textsc{Cms}
\cite{cms17_searc_third_gener_scalar_leptoq, cms17_searc_heavy_neutr_or_third,
  cms16_searc_heavy_major_neutr_e, cms15_searc_heavy_major_neutr}.  These were
performed in the context of other seesaw models, though in all cases, they
selected for events with the topology shown in
\cref{subfig:neutrino_search_topology}.  Often, searches look for same-sign
leptons and also different-flavour lepton as the \textsc{sm} backgrounds are low
in both cases.  Unfortunately, the cross section \(\sigma(pp \to \ell N)\) is
generally too small as it is suppressed by the neutrino mixing.  As discussed in
\cref{subsec:neutrino_masses_and_mixing}, it is possible to have quite a
considerable mixing though this is bounded by precision lepton flavour and
electroweak observables.  For regions which are not already excluded by
\textsc{lep} and the precision observables, the production cross section at
\SI{13}{\TeV} reaches at most \SI{e-3}{\pb} which is too small to be observed at
the \textsc{lhc} over the \textsc{sm} backgrounds.

On the other hand, the production of the exotic lepton doublet can be much
larger as it is generally not suppressed by the neutrino mixing.  The topology
of the event in this case is more complex, which would consist of two \(W\)
bosons, one lepton, and missing \(E_{T}\) as shown in
\cref{subfig:exotic_lepton_search_topology}.  The \(W\) bosons could either
decay hadronically which would facilitate the event reconstruction, or they
could decay to different-flavour leptons in order to avoid \textsc{sm}
background though at the cost of having several neutrinos in the final
state. Nevertheless, the production cross section for this topology is of order
\SI{0.1}{\pb} at best in regions which are not excluded by \textsc{lep}, and
falls off very rapidly as the masses of the leptons are increased.

Although this model does not present a clear way to detect the heavy neutrinos
directly, the phenomenology presented by the remaining exotic particle content
can provide indirect bounds on the heavy neutrino masses.  In particular, we
have not been imposing that the Yukawa couplings unify at the \textsc{gut}
scale, though a complete model must evidently do so.  If we require that
\(y_{1d} = y_{1\ell}\) at the \textsc{gut} scale, the bounds on the exotic
quarks discussed in \cref{subsec:exotic_down_quarks} translate to a bound of
\(y_{1\ell} > \num{0.040}\) (\(m_{r}, m_{N_{1}} > \SI{650}{\GeV}\)) in the
collider-stable case, and \(y_{1\ell} > \num{0.024}\) (\(m_{r}, m_{N_{1}} >
\SI{390}{\GeV}\)) in the prompt-decay case.

\subsection{Higgs Sector}
\label{subsec:higgs_sector}

The two Higgs doublets contained within the scalar sector correspond to the
well-studied type-\textsc{ii} \textsc{2hdm} (see
Refs.~\cite{bhattacharyya16_scalar_sector_two_higgs_doubl_model,
  branco12_theor_phenom_two_higgs_doubl_model} for recent reviews) in which one
gauge eigenstate couples to up-type quarks (and neutrinos), and the other
couples to down-type quarks and electrons.  The terms of the scalar potential
concerning the Higgs doublets are
\begin{equation}
  \label{eq:2hdm_potential}
  \begin{split}
    \MoveEqLeft[1] \calV = - \mu_{u}^{2} H_{u}^{\dagger} H_{u} - \mu_{d}^{2} H_{d}^{\dagger} H_{d}
    + \lambda_{u} (H_{u}^{\dagger} H_{u})^{2}
    + \lambda_{d} (H_{d}^{\dagger} H_{d})^{2} \\
    &+ \lambda_{ud} (H_{u}^{\dagger} H_{u}) (H_{d}^{\dagger} H_{d})  
    + \lambda_{ud}' (H_{d}^{\dagger} H_{u}) (H_{u}^{\dagger} H_{d})  
    - \left[ \kappa \Phi_{1} H_{u} H_{d} \phc \right]
  \end{split}
\end{equation}
After \(\Phi_{1}\) gains a \textsc{vev}, the last term generates the term
\(\mu_{ud}^{2} \defeq \kappa v_{1} / \sqrt{2}\) (also referred to as \(m_{12}\)
in the literature).  As discussed in \cref{subsec:exotic_Z_bosons}, there
already exists a lower limit on \(v_{1}\) of \(\sim \SI{23}{\TeV}\), thus
\(\mu_{ud}\) can easily be large.  Having said this, the limit \(\kappa \to 0\)
is technically natural as it breaks an accidental global \Uone{}
symmetry,\footnote{Under this accidental global \Uone{} symmetry, the charges of
  \(\Phi_{1}\), \(\Phi_{2}\), \(\chi_{x, \overline 5}\) and \(\chi_{1}\) are
  \(-2\), \(1\), \(1\) and \(-1\) respectively: all other fields remain
  uncharged.  The \textsc{d5} term in \cref{eq:d5_terms_no_phi3} is forbidden
  under this symmetry and the \textsc{d5} terms in \cref{eq:d5_terms_with_phi3}
  are allowed provided \(\Phi_{3}\) has charge \(-1\).} and thus it is not
regenerated through radiative corrections and can be small.  The
\((H_{u}^{\dagger} H_{d})^{2}\) term present in conventional \textsc{2hdm}
models is forbidden here as it is not gauge invariant under the additional
\Uone{} gauge groups.  As a result, the only term which is capable of
introducing \(CP\) violation is \(\kappa\), though we will be taking this to be
real.

The pseudoscalar and charged scalar squared masses are
\begin{subequations}
  \begin{align}
    \label{eq:pseudoscalar_mass}
    m_{A}^{2} &= \frac{\kappa v_{\SM}^{2} \sin 2\beta}{2 \sqrt{2} v_{1}} + \frac{\sqrt{2} \kappa v_{1}}{\sin 2\beta}, \\
    \label{eq:charged_scalar_mass}
    m_{H^{+}}^{2} &= \frac{\lambda_{ud}'}{2} v_{\SM}^{2} + \frac{\sqrt{2} \kappa v_{1}}{\sin 2\beta}.
  \end{align}
\end{subequations}
Additionally, there are two neutral scalars that arise from \textsc{2hdm}: \(h\)
and \(H\).  One of them is the \textsc{sm}-like Higgs with \(m_{h} =
\SI{125}{\GeV}\) and the second scalar will in general be heavier.  Their
squared masses are
\begin{subequations}
  \label{eq:2hdm_scalar_mass}
  \begin{align}
    m_{h}^{2} &\approx \frac{v_{\SM}^{2}}{2} \Big[
                \lambda_{d} \cos^{4} \beta
                + \lambda_{u} \sin^{4} \beta
                + \lambda_{ud} \sin^{2} 2\beta \Big], \\
    \begin{split}
      m_{H}^{2} &\approx \frac{\sqrt{2} \kappa v_{1}}{\sin 2 \beta}
      + \frac{v_{\SM}^{2}}{8} \Big[
      \lambda_{d} \sin^{4} 2 \beta
      + \lambda_{u} \sin^{4} 2 \beta
      - 4 \lambda_{ud} \sin^{2} 2\beta \Big],
    \end{split}
  \end{align}
\end{subequations}
assuming that the mixing with the \(CP\)-even components of \(\Phi_{1,2}\) can
be neglected.  With the exception of very small values for \(\kappa v_{1}\) and
\(\tan \beta\), it will in general be the case that \(m_{H} \approx m_{H^{+}}
\approx m_{A}\).

The rotation of the Higgs doublets from the gauge basis to the mass basis is
described by the angle \(\alpha\), while the rotation from the gauge basis to
the \emph{Higgs basis} is described by \(\beta\) (which has been defined
earlier).  The Higgs basis is defined such that only one (the \textsc{sm} Higgs)
contains the \textsc{vev}.  The scenario in which the mass eigenstates line up
with the Higgs basis is called the \emph{alignment
  limit}\footnote{Ref.~\cite{branco12_theor_phenom_two_higgs_doubl_model} refers
  to this as the decoupling limit (in which the masses of the exotic scalars are
  much heavier), though as pointed out in Ref.~\cite{Dev:2014yca}, it is
  possible to achieve alignment without decoupling.} and corresponds to
\(\cos(\alpha - \beta) \to 0\) and deviations from this limit are strongly
disfavoured.  In this model, the large \textsc{vev} of \(\Phi_{1}\) helps to
ensure the alignment limit, with
\begin{equation}
  \label{eq:higgs_alignment}
  \cos(\alpha - \beta) \approx \frac{v_{\SM}^{2}}{\sqrt{2} \kappa v_{1}} \frac{\lambda_{u} - 2 \lambda_{ud}}{\tan^{2} \beta}.
\end{equation}

The constraints on \textsc{2hdm} models were most recently collated in
Ref.~\cite{arbey17_status_charg_higgs_boson_two}.  One of the more stringent
constraints relevant to type-\textsc{ii} models arises from \(b \to s\)
transitions which are facilitated by the charged Higgs.  Specifically, the
latest constraints require that \(m_{H^{+}} > \SI{600}{\GeV}\) for \(\tan \beta
> 1\), and only increase with \(\tan \beta < 1\).

\begin{figure}
  \centering
  \tikzsetnextfilename{kappa_tanbeta}
  \begin{tikzpicture}
    \begin{semilogyaxis}[
        width=\figwidth\linewidth,
        height=\figheight\linewidth,
        xlabel={\(\kappa v_{1}\) [\si{\GeV^2}]},
        ylabel={\(\tan \beta\)},
        ytick={1, 2, 5, 10, 20},
        minor ytick={3, 4, 6, 7, 8, 9, 12, 14, 16, 18, 22, 24, 26, 28, 30},
        yticklabels={1, 2, 5, 10, 20},
        view={0}{90},
        point meta rel=per plot,
        unbounded coords=jump,
        table/col sep=comma,
        table/x=kappa-v1,
        table/y=tanbeta,
      ]

      \addplot3[
        Dark2-A,
        line legend,
        contour gnuplot={
          levels={0.5},
          draw color={Dark2-A},
          labels=false,
        },
      ] table [z=pseudoscalar] {paper_data/2hdm.csv} ;
      \addlegendentry{{\(pp \to A \to \tau\tau\)}};

      \addplot3[
        Dark2-B,
        line legend,
        contour gnuplot={
          levels={0.5},
          draw color={Dark2-B},
          labels=false,
        },
      ] table [z=charged] {paper_data/2hdm.csv};
      \addlegendentry{{\(b \to sX\)}};
    \end{semilogyaxis}
  \end{tikzpicture}
  \caption{Bounds on the \textsc{2hdm} sector of the model, presented on the
    parameters \(\tan \beta\) and \(\kappa v_{1}\).  The lower left area of the
    plot is excluded.  The bounds were recast from
    Ref.~\cite{arbey17_status_charg_higgs_boson_two}.}
  \label{fig:2hdm_constraints}
\end{figure}
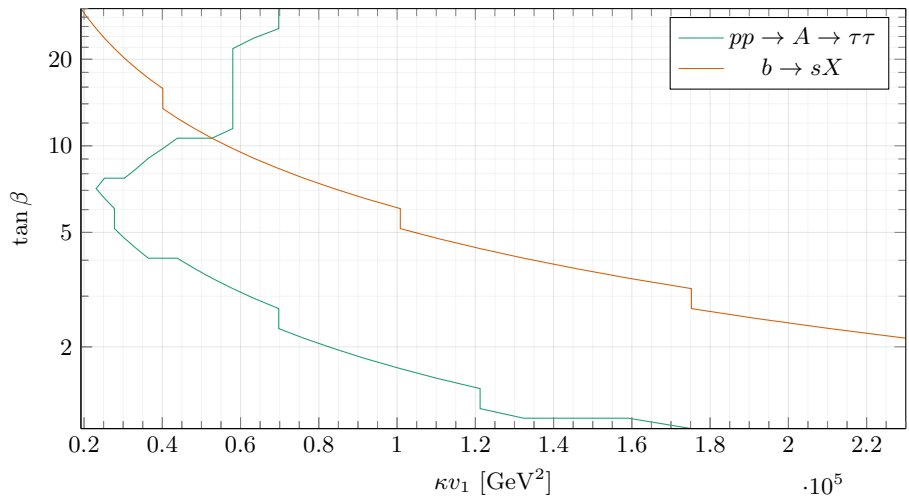

In the context of collider searches, the pseudoscalar's coupling to the bottom
quarks and \(\tau\) leptons is enhanced by larger values of \(\tan \beta\). This
allows for the pseudoscalar to be produced through a loop of \(b\) quarks and
subsequently decay to a pair of \(\tau\) leptons providing constraints up to
\SI{1}{\TeV} in the large-\(\tan \beta\) regime.  The most restrictive
constraints from Ref.~\cite{arbey17_status_charg_higgs_boson_two} have been
recast on the \(\tan \beta\)--\(\kappa v_{1}\) parameter space and are shown in
\cref{fig:2hdm_constraints}, assuming the benchmark configuration of
\textsc{vev}s.  Values of \(\kappa v_{1}\) smaller than \SI{60e3}{\GeV^2} are
generally excluded and larger values of \(\tan \beta\) are favoured.

\section{Conclusion}

In this paper, we have explored the phenomenology of the model presented in
Ref.~\cite{cai16_tev_scale_pseud_dirac_seesaw}.  The primary motivation for this
model---the realization of a seesaw mechanism at mass scales testable at the
\textsc{lhc}---was achieved without recourse to exceptionally small Yukawa
couplings.  The additional neutral fermions required to generate the seesaw
mechanism are motivated by unification into \Esix{} which, in order to complete
the \(\bm{27}\) irrep, requires the existence of additional charged leptons and
isosinglet quarks as well as two \(Z'\) bosons.  The masses of the exotic
fermions are generated by the \textsc{vev}s \(v_{1,2}\) of two scalar singlets
\(\Phi_{1,2}\), and the model also presents a type-\textsc{ii} two-Higgs-doublet
model.

A significant constraint on this model arises from searches for \(Z'\) bosons
which restrict the lightest \(Z'\) mass to be above \SI{2.8}{\TeV}.  As the
\(Z'\) mass is generated by \(v_{1,2}\) and the gauge couplings are fixed by
requiring that they unify at the \textsc{gut} scale, the \(Z'\) bound imposes
that \(v_{1}\) and \(v_{2}\) be larger than \SI{18}{\TeV} and \SI{15}{\TeV}
respectively.  This will, in general, lead to large masses for the scalar
singlets present in this model.

A small amount of mixing between the \textsc{sm} and exotic down-type quarks is
generated through a dimension-five term which allows the exotic quarks to decay
before \textsc{bbn}, but results in collider-stable quarks.  As the mixing is
reduced with larger mass separations, \textsc{bbn} imposes an upper bound on the
exotic quark masses of \SI{3}{\TeV} for the benchmark configuration of
\textsc{vev}s investigated in this model, and collider searches for long-lived
particles restrict the masses to be above \SI{1.3}{\TeV}.  In order to
facilitate the decay of exotic quarks, the authors of
Ref.~\cite{cai16_tev_scale_pseud_dirac_seesaw} introduced another scalar which
gains a \textsc{vev}.  This can allow the exotic quarks to decay promptly if
produced at the \textsc{lhc} in which case searches place a lower bound on their
masses of \SI{810}{\GeV}.

Searches for the exotic leptons at the \textsc{lhc} are unfortunately not
feasible as the production cross section is generally far too small.  In
particular, if a prompt \textsc{sm} lepton is required, the process is greatly
suppressed by the neutrino mixing.  Even the direct production of the exotic
charged leptons is generally too small to be seen over the background.  As a
result, the bound on the leptons remains at \SI{102}{\GeV} from searches at
\textsc{lep}.

Finally, the model also contains a type-\textsc{ii} \textsc{2hdm} with the
addition of a \(\kappa \Phi_{1} H_{u} H_{d}\) interaction in the potential.  The
large \textsc{vev} of \(\Phi_{1}\) helps enforce the alignment limit in the
model thereby avoiding most constraints that arise from the \textsc{2hdm}.
Values of \(\kappa v_{1}\) below \SI{60e3}{\GeV^2} are excluded, and larger
values of \(\tan \beta\) are favoured.


\begin{acknowledgements}
  J.\,P.\,E. thanks Jackson Clarke for the valuable discussions and guidance
  during this research.  J.\,P.\,E. and R.\,R.\,V. also thank Jackson Clarke and Yi
  Cai for carefully reading and providing valuable feedback on an earlier draft
  of this paper.  This work was supported in part by the Australian Research
  Council and the Commonwealth of Australia.  All Feynman diagrams were drawn
  using Ti\emph{k}Z-Feynman \cite{ellis16_tikz_feynm_diagrams_tikz}.
\end{acknowledgements}

\appendix

\section{\texorpdfstring{\(B\) Partial Widths}{B Partial Widths}}
\label{sec:B_partial_widths}

The mass matrix of the \(b\)--\(B\) quark is shown in \cref{eq:b-B_mass_matrix},
and can be diagonalized using the two unitary matrices \(U_{\scL}\) and
\(U_{\scR}\) as shown in \cref{eq:b-B_diagonal_mass_matrix} where the (un)primed
fields denote (gauge) mass eigenstates.  We are assuming no mixing between
generations so that it is sufficient to deal with a \(2 \times 2\) mass
matrix. When \(\Phi_{3}\) is present, the off-diagonal term in
\cref{eq:b-B_mass_matrix} becomes \((v_{1} v_{2} + v_{1} v_{3}) / 2 \Lambda\)
and for simplicity, we will define \(f \defeq (v_{1} v_{2} + v_{1} v_{3}) / 2
\Lambda\).  By taking \(v_{3} \to 0\), we recover the scenario where
\(\Phi_{3}\) is absent.

To leading order in \(f\) and \(m_{b} / m_{B}\), the two matrices rotating the
gauge eigenstates to the mass basis as defined in
\cref{eq:b-B_diagonal_mass_matrix} are
\begin{subequations}
  \begin{align}
    \label{eq:B_rotation_matrices}
    U_{\scL} &= \begin{pmatrix}
      1 & f \frac{m_{b}}{m_{B}^{2}} \\
      - f \frac{m_{b}}{m_{B}^{2}} & 1
    \end{pmatrix}, &
    U_{\scR} &= \begin{pmatrix}
      1 & f \frac{1}{m_{B}} \\
      -f \frac{1}{m_{B}} & 1
    \end{pmatrix}.
  \end{align}
\end{subequations}

In the gauge basis, the \textsc{sm} and exotic bottom quark couplings to the
\(W\), \(Z\) gauge bosons and the Higgs are,
\begin{subequations}
  \begin{align}
    \label{eq:Z_couplings}
      \calL_{Z} &=
      \begin{pmatrix}
        \overline b_{\scL} & \overline B_{\scL}
      \end{pmatrix}
      \slashed Z
      \begin{pmatrix}
        g^{Z}_{\scL, b} & 0 \\
        0 & g^{Z}_{B}
      \end{pmatrix}
      \begin{pmatrix}
        b_{\scL} \\ B_{\scL}
      \end{pmatrix}
      +
      \begin{pmatrix}
        \overline b_{\scR} & \overline B_{\scR}
      \end{pmatrix}
      \slashed Z
      \begin{pmatrix}
        g^{Z}_{\scR, b} & 0 \\
        0 & g^{Z}_{B}
      \end{pmatrix}
      \begin{pmatrix}
        b_{\scR} \\ B_{\scR}
      \end{pmatrix},
    \\
    \label{eq:W_couplings}
    \calL_{W} &= \frac{g_{2}}{\sqrt{2}}
    \begin{pmatrix}
      \overline b_{\scL} & \overline B_{\scL}
    \end{pmatrix}
    \slashed W
    \begin{pmatrix}
      1 & 0 \\
      0 & 0
    \end{pmatrix}
    \begin{pmatrix}
      b_{\scL} \\ B_{\scL}
    \end{pmatrix},
    \\
    \calL_{h} &= \frac{h}{\sqrt{2}}
    \begin{pmatrix}
      \overline b_{\scL} & \overline B_{\scL}
    \end{pmatrix}
    \begin{pmatrix}
      y_{b} & \frac{v_{3}}{\sqrt{2} \Lambda} \\
      0 & 0
    \end{pmatrix}
    \begin{pmatrix}
      b_{\scR} \\ B_{\scR}
    \end{pmatrix},
  \end{align}
\end{subequations}
where the couplings to the \(Z\) boson are
\begin{subequations}
  \begin{align}
    g^{Z}_{\scL,b} &= \frac{1}{60} \mathcal{Z}_{i,2} \begin{pmatrix}
      -10 g_{1} \\ 30 g_{2} \\ 3 \sqrt{5} g_{4} \\ 5 g_{5}
    \end{pmatrix}_{i}, &
    g^{Z}_{\scR,b} &= \frac{1}{60} \mathcal{Z}_{i,2} \begin{pmatrix}
      20 g_{1} \\ 0 \\ 9 \sqrt{5} g_{4} \\ -5 g_{5}
    \end{pmatrix}_{i}, &
    g^{Z}_{B} &= \frac{1}{30} \mathcal{Z}_{i,2} \begin{pmatrix}
      10 g_{1} \\ 0 \\ -3 \sqrt{5} g_{4} \\ -5 g_{5}
    \end{pmatrix}_{i}.
  \end{align}
\end{subequations}
in which \(\mathcal{Z}\) is the matrix that rotates the neutral gauge bosons
into their mass eigenstates [see \cref{eq:gauge_boson_mass_matrix} for the mass
matrix] defined such that
\begin{equation}
  \begin{pmatrix}
    \gamma \\ Z \\ Z' \\ Z''
  \end{pmatrix}
  =
  \mathcal{Z}
  \begin{pmatrix}
    A_{\scY} \\ B_{\SU{2}} \\ A_{\chi} \\ A_{\psi},
  \end{pmatrix}
\end{equation}
\(g_{i}\) are the gauge couplings, and \(y_{b}\) is the bottom-quark Yukawa. The
interactions in the mass basis are then obtained by appropriately rotating the
left and right components with \(U_{\scL,\scR}\), after which the resulting
partial widths of the exotic quarks are
\begin{subequations}
  \begin{align}
    \label{eq:B_to_Wt_partial_width}
      \Gamma(B' \to Wt)
      &= \frac{\lambda^{\frac{1}{2}}(m_{B}^{2}, m_{W}^{2}, m_{t}^{2})}{16 \pi}
      \frac{m_{B}}{m_{W}^{2}} \abs*{\hat{g}^{W}}^{2}
      \Bigl[ 1 + r_{W}^{2} - 2 r_{t}^{2} - 2 r_{W}^{4} + r_{t}^{4} + r_{W}^{2} r_{t}^{2} \Bigr] \\
    \label{eq:B_to_Zb_partial_width}
    \begin{split}
      \Gamma(B' \to Z b')
      &= \frac{\lambda^{\frac{1}{2}}(m_{B}^{2}, m_{Z}^{2}, m_{b}^{2})}{16 \pi} \frac{m_{B}}{m_{Z}^{2}}
      \Bigl[ \\
      &\qquad \left( \abs*{\hat{g}^{Z}_{\scL}}^{2} + \abs*{\hat{g}^{Z}_{\scR}}^{2} \right)
      \left( 1 + r_{Z}^{2} - 2 r_{b}^{2} - 2 r_{Z}^{4} + r_{b}^{4} + r_{Z}^{2} r_{b}^{2} \right) \\
      &\qquad - 12 \Re[\hat{g}^{Z}_{\scL} \hat{g}^{Z}_{\scR}] r_{b} r_{Z}^{2}
      \Bigr]
    \end{split} \\
    \label{eq:B_to_hb_partial_width}
    \begin{split}
      \Gamma(B' \to h b')
      &= \frac{\lambda^{\frac{1}{2}}(m_{B}^{2}, m_{h}^{2}, {m_{b}'}^{2})}{16 \pi} \frac{1}{m_{B}}
      \Bigl[
      \left( \abs*{\hat{y}_{1}}^{2} + \abs*{\hat{y}_{2}}^{2} \right) \left( 1 - r_{h}^{2} + r_{b}^{2} \right)
      + 4 \Re[ \hat{y}_{1} \hat{y}_{2} ] r_{b}
      \Bigr]
    \end{split}
  \end{align}
\end{subequations}
where \(\lambda(x, y, z)\) is the K\"all\'en function
\begin{equation*}
  \lambda(x, y, z) \defeq x^2 + y^2 + z^2 - 2xy - 2xz - 2yz,
\end{equation*}
\(r_{i} \defeq m_{i} / m_{B}\), and the couplings with circumflexes take into
account the rotation to the mass basis.  Note that as the corrections to the
masses are extremely small [see \cref{eq:b_mass_corrections}], the distinction
between \(m_{b,B}\) and \(m_{b,B}'\) is omitted in the above partial
widths. Explicitly, the couplings appearing in
\cref{eq:B_to_Wt_partial_width,eq:B_to_Zb_partial_width,eq:B_to_hb_partial_width}
are
\begin{subequations}
  \begin{align}
    \hat{g}^{W} &= f \frac{m_{b}}{m_{B}^{2}} \frac{g_{2}}{\sqrt{2}} , \\
    \hat{g}^{Z}_{\scL} &= 2 f \frac{m_{b}}{m_{B}^{2}} \left( g^{Z}_{B} -  g^{Z}_{\scL,b} \right), &
    \hat{g}^{Z}_{\scR} &= 2 f \frac{1}{m_{B}} \left( g^{Z}_{\scR,b} - g^{Z}_{B} \right), \\
    \hat{y}_{1} &= - f \frac{1}{m_{B}} \frac{y_{b}}{\sqrt{2}} - f^{2} \frac{1}{m_{B}} \frac{m_{b}}{m_{B}^{2}} \frac{v_{3}}{2 \Lambda}, &
    \hat{y}_{2} &= - f \frac{m_{b}}{m_{B}^{2}} \frac{y_{b}}{\sqrt{2}} + \frac{v_{3}}{2 \Lambda}.
  \end{align}
\end{subequations}


\bibliography{references}

\end{document}

%% file: pgfplots.default.tex
\pgfplotsset{
  unbounded coords=discard,
  filter discard warning=false,
  every axis/.append style={
    minor tick num=3,
    grid=both,
    major grid style={
      color=black,
      opacity=0.1,
    },
    minor grid style={
      color=black,
      opacity=0.05,
    },
  },
  colormap/magma/.style={%
    /pgfplots/colormap={magma}{%
      rgb=(0.001462, 0.000466, 0.013866)
      rgb=(0.035520, 0.028397, 0.125209)
      rgb=(0.102815, 0.063010, 0.257854)
      rgb=(0.191460, 0.064818, 0.396152)
      rgb=(0.291366, 0.064553, 0.475462)
      rgb=(0.384299, 0.097855, 0.501002)
      rgb=(0.475780, 0.134577, 0.507921)
      rgb=(0.569172, 0.167454, 0.504105)
      rgb=(0.664915, 0.198075, 0.488836)
      rgb=(0.761077, 0.231214, 0.460162)
      rgb=(0.852126, 0.276106, 0.418573)
      rgb=(0.925937, 0.346844, 0.374959)
      rgb=(0.969680, 0.446936, 0.360311)
      rgb=(0.989363, 0.557873, 0.391671)
      rgb=(0.996580, 0.668256, 0.456192)
      rgb=(0.996727, 0.776795, 0.541039)
      rgb=(0.992440, 0.884330, 0.640099)
      rgb=(0.987053, 0.991438, 0.749504)
    },
  },
  colormap/inferno/.style={%
    /pgfplots/colormap={inferno}{%
      rgb=(0.001462, 0.000466, 0.013866)
      rgb=(0.037668, 0.025921, 0.132232)
      rgb=(0.116656, 0.047574, 0.272321)
      rgb=(0.217949, 0.036615, 0.383522)
      rgb=(0.316282, 0.053490, 0.425116)
      rgb=(0.410113, 0.087896, 0.433098)
      rgb=(0.503493, 0.121575, 0.423356)
      rgb=(0.596940, 0.154848, 0.398125)
      rgb=(0.688653, 0.192239, 0.357603)
      rgb=(0.775059, 0.239667, 0.303526)
      rgb=(0.851384, 0.302260, 0.239636)
      rgb=(0.912966, 0.381636, 0.169755)
      rgb=(0.956852, 0.475356, 0.094695)
      rgb=(0.981895, 0.579392, 0.026250)
      rgb=(0.987464, 0.690366, 0.079990)
      rgb=(0.973088, 0.805409, 0.216877)
      rgb=(0.947594, 0.917399, 0.410665)
      rgb=(0.988362, 0.998364, 0.644924)
    },
  },
  colormap/plasma/.style={%
    /pgfplots/colormap={plasma}{%
      rgb=(0.050383, 0.029803, 0.527975)
      rgb=(0.186213, 0.018803, 0.587228)
      rgb=(0.287076, 0.010855, 0.627295)
      rgb=(0.381047, 0.001814, 0.653068)
      rgb=(0.471457, 0.005678, 0.659897)
      rgb=(0.557243, 0.047331, 0.643443)
      rgb=(0.636008, 0.112092, 0.605205)
      rgb=(0.706178, 0.178437, 0.553657)
      rgb=(0.768090, 0.244817, 0.498465)
      rgb=(0.823132, 0.311261, 0.444806)
      rgb=(0.872303, 0.378774, 0.393355)
      rgb=(0.915471, 0.448807, 0.342890)
      rgb=(0.951344, 0.522850, 0.292275)
      rgb=(0.977856, 0.602051, 0.241387)
      rgb=(0.992541, 0.687030, 0.192170)
      rgb=(0.992505, 0.777967, 0.152855)
      rgb=(0.974443, 0.874622, 0.144061)
      rgb=(0.940015, 0.975158, 0.131326)
    },
  },
  colormap/jpeBlWhBl/.style={%
    /pgfplots/colormap={jpeReWhBl}{%
      rgb=(0.094957, 0.211745, 0.949475)
      rgb=(0.149676, 0.288182, 0.889340)
      rgb=(0.210373, 0.340936, 0.846637)
      rgb=(0.268241, 0.383971, 0.815988)
      rgb=(0.322709, 0.422216, 0.794156)
      rgb=(0.374335, 0.457967, 0.779078)
      rgb=(0.423724, 0.492464, 0.769388)
      rgb=(0.471369, 0.526452, 0.764155)
      rgb=(0.517653, 0.560415, 0.762732)
      rgb=(0.562866, 0.594688, 0.764671)
      rgb=(0.607234, 0.629520, 0.769662)
      rgb=(0.650934, 0.665108, 0.777508)
      rgb=(0.694107, 0.701609, 0.788103)
      rgb=(0.736875, 0.739156, 0.801431)
      rgb=(0.779363, 0.777851, 0.817590)
      rgb=(0.821732, 0.817752, 0.836891)
      rgb=(0.863910, 0.858922, 0.860421)
      rgb=(0.902313, 0.903243, 0.885111)
      rgb=(0.879049, 0.855058, 0.839087)
      rgb=(0.860371, 0.806330, 0.794603)
      rgb=(0.843643, 0.758010, 0.750027)
      rgb=(0.828754, 0.709912, 0.704933)
      rgb=(0.815142, 0.662021, 0.659169)
      rgb=(0.802302, 0.614305, 0.612550)
      rgb=(0.789774, 0.566724, 0.564903)
      rgb=(0.777122, 0.519227, 0.516091)
      rgb=(0.763930, 0.471751, 0.466024)
      rgb=(0.749803, 0.424208, 0.414668)
      rgb=(0.734369, 0.376470, 0.362052)
      rgb=(0.717286, 0.328322, 0.308250)
      rgb=(0.698243, 0.279394, 0.253349)
      rgb=(0.676967, 0.228987, 0.197339)
      rgb=(0.653216, 0.175627, 0.139788)
      rgb=(0.626782, 0.115406, 0.078551)
      rgb=(0.597480, 0.029268, 0.008874)
    },
  },
  colormap/jpeWhRe/.style={%
    /pgfplots/colormap={jpeWhRe}{%
      rgb=(0.902313, 0.903243, 0.885111)
      rgb=(0.879049, 0.855058, 0.839087)
      rgb=(0.860371, 0.806330, 0.794603)
      rgb=(0.843643, 0.758010, 0.750027)
      rgb=(0.828754, 0.709912, 0.704933)
      rgb=(0.815142, 0.662021, 0.659169)
      rgb=(0.802302, 0.614305, 0.612550)
      rgb=(0.789774, 0.566724, 0.564903)
      rgb=(0.777122, 0.519227, 0.516091)
      rgb=(0.763930, 0.471751, 0.466024)
      rgb=(0.749803, 0.424208, 0.414668)
      rgb=(0.734369, 0.376470, 0.362052)
      rgb=(0.717286, 0.328322, 0.308250)
      rgb=(0.698243, 0.279394, 0.253349)
      rgb=(0.676967, 0.228987, 0.197339)
      rgb=(0.653216, 0.175627, 0.139788)
      rgb=(0.626782, 0.115406, 0.078551)
      rgb=(0.597480, 0.029268, 0.008874)
    },
  },
  colormap/viridis,
}
